\documentclass[%
 reprint,
 amsmath,amssymb,
 aps,
]{revtex4-2}

\usepackage{graphicx}% Include figure files
\usepackage{dcolumn}% Align table columns on decimal point
\usepackage{bm}% bold math
\usepackage{hyperref}% add hypertext capabilities
\RequirePackage[all]{hypcap}
\RequirePackage[capitalise,nameinlink]{cleveref}
\usepackage[english]{babel}
\usepackage{svg}
\usepackage{bookmark}
\newcommand{\wh}{\omega_{1/2}}
\newcommand{\dV}{\delta V}
\newcommand{\dP}{\delta\phi}
\newcommand{\dw}{\delta\omega}
\newcommand{\Dw}{\Delta\omega}
\newcommand{\wmu}{\Omega_\mu}
\newcommand{\dI}{\delta I}
\newcommand{\dQ}{\delta Q}
\begin{document}

\preprint{APS/123-QED}

\title{The Ponderomotive Effects of Narrow-band,\texorpdfstring{\\}{ }Superconducting Resonators in Open and Closed Loop}% Force line breaks with \\
\thanks{This material is based upon work supported by the U.S. Department of Energy, Office of Science, Office of Workforce Development for Teachers and Scientists, Office of Science Graduate Student Research (SCGSR) program. The SCGSR program is administered by the Oak Ridge Institute for Science and Education for the DOE under contract number DE‐SC0014664}%

\author{Jacob Brown}

 \email{brownjac@frib.msu.edu}
\affiliation{%
 Facility for Rare Isotope Beams (FRIB)\\
 640 S. Shaw Ln., East Lansing, MI 48824 USA
}%
\affiliation{Department of Physics and Astronomy,\\
Michigan State University,\\
567 Wilson Rd., East Lansing, MI 48824 USA}
\affiliation{
 Fermi National Accelerator Laboratory (FNAL),\\
 Kirk and Pine St., Batavia, IL 60510 USA% with \\
}
%\altaffiliation[Also at ]{Fermi National Accelerator Laboratory, Kirk and Pine St., Batavia, IL 60510}%Lines break automatically or can be forced with \\

\author{Alexander Sukhanov}
\email{ais@fnal.gov}
\author{Crispin Contreras-Martinez}
\author{Vyacheslav Yakovlev}

 %\homepage{http://www.Second.institution.edu/~Charlie.Author}
\affiliation{
 Fermi National Accelerator Laboratory (FNAL),\\
 Kirk and Pine St., Batavia, IL 60510 USA % with \\
}%

\author{Sang-hoon Kim}
\author{Shen Zhao}
\author{Ting Xu}
\altaffiliation[Also at ]{the Department of Physics and Astronomy,\\
Michigan State University.}
\author{Walter Hartung}
\author{Wei Chang}
\affiliation{%
 Facility for Rare Isotope Beams (FRIB)\\
 640 S. Shaw Ln., East Lansing, MI 48824 USA
}%

%\collaboration{CLEO Collaboration}%\noaffiliation

\date{\today}% It is always \today, today,
             %  but any date may be explicitly specified

\begin{abstract}
In this work, we present measurements of ponderomotive instabilities in narrow-band, coaxial resonators in open and closed control loop systems. We show an analytical scheme we will use in future studies to examine the dependency of open loop stability on mechanical parameters. Analytical and simulation models are used to predict the onset of the oscillatory instability in half-wave resonators with active disturbance rejection control for amplitude and phase stabilization. We demonstrate that for high amplitude controller bandwidths, ponderomotive oscillations can couple with the controller frequency response via higher harmonics and lower the threshold for the oscillatory instability. We reaffirm the superiority of in-phase/quadrature ($I/Q$ ) component control in regards to preventing the oscillatory instability, and show how the phase controller can amplify the crosstalk of disturbances in amplitude and phase control. In cases with large disturbances, this effect can lead to instabilities not present in $I/Q$ control.
%\begin{description}
%\item[Usage]
%Secondary publications and information retrieval purposes.
%\item[Structure]
%You may use the \texttt{description} environment to structure your abstract;
%use the optional argument of the \verb+\item+ command to give the category of each item. 
%\end{description}
\end{abstract}

%\keywords{Suggested keywords}%Use showkeys class option if keyword
                              %display desired
\maketitle

%\tableofcontents

\section{\label{sec:level1}Introduction}
\subsection{Coaxial Resonators}
Superconducting coaxial resonators have been successfully utilized in low-velocity acceleration of charged particles \cite{FORTUNA1990253, osti_5991052, ReA, spiral2}. Current and next generation linear, heavy-ion and hadron accelerators (linacs) in America have and will utilize these structures in their medium velocity acceleration scheme. The Facility for Rare Isotope Beams, which began user operations in 2022 \cite{mpla37:2230006, osti_og}, utilizes four varieties of coaxial resonators across 46 cryomodules \cite{Xu:SRF2017-TUXAA03}. The medium-velocity acceleration is handled by two varieties of half-wave resonators (HWRs) operating at 322~MHz: one optimized for $\beta=\frac{v}{c}=0.29$ and the other for $\beta=0.53$.

The Proton Improvement Plan (PIP-II) at Fermilab will power the next generation of neutrino science in the United States. The 800~MeV superconducting $H^-$ linac contains two varieties of single-spoke resonators (SSRs) operating at 325~MHz. The first variety, dubbed SSR1, is optimized for $\beta=0.22$. The second type, SSR2, is optimized for $\beta=0.47$ \cite{pip2}. 

\begin{figure}
    \centering
    \includegraphics[width=\linewidth]{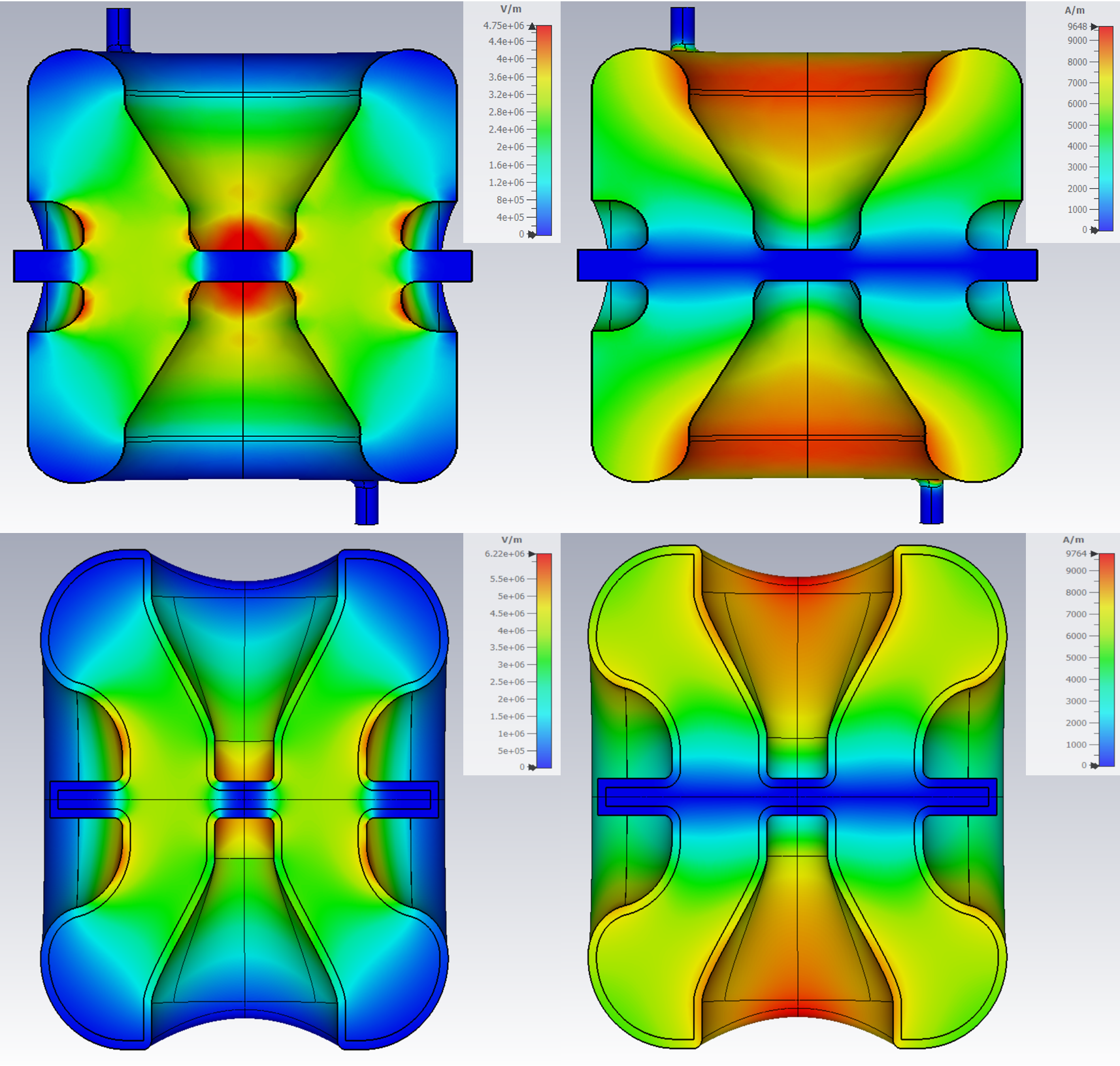}
    \caption{CST simulations for the FRIB $\beta=0.53$ HWR's electric and magnetic fields respectively (top, left to right), and PIP-II SSR2 electric and magnetic fields (bottom, left to right). Fields are normalized to an internal energy of 1~J.}
    \label{fig:fields}
\end{figure}

Geometrically, there is not much difference between SSRs and HWRs; the main difference lies in shaping of the cavity near the short planes and the beam cups. We present the fundamental eigenmode fields for the $\beta=0.53$ HWR and SSR2 in \cref{fig:fields}. These 4 cavities also share similar electrical parameters. We list the relevant parameters in \cref{tab:cav-params}: the design frequency $f_0$, the cavity effective length $l_{eff}$, the max accelerating electric field ($E_{acc}$), the loaded quality factor $Q_L$, and resulting loaded half-bandwidth $f_{1/2}$.

\begin{table}[]
    \centering
    \begin{tabular}{c|c|c|c|c}
    \hline
    Param. [Units]  &  $\beta=0.29$ & $\beta=0.53$ & $\beta=0.22$ & $\beta=0.47$\\
    \hline
    $f_0$ [MHz]     & 322 & 322 & 325 & 325\\
    $l_{eff}$ [m] & 0.27 & 0.503 & 0.205 & 0.438\\
    Max $E_{acc}$ [MV/m] & 7.7 & 7.4 & 10 & 11.4\\
    $Q_L$ [NA] & $5\times10^6$ & $1\times10^7$ & $3\times10^6$ & $5\times10^6$\\
    $f_{1/2}$ [Hz] & 28.5 & 16.1 & 54.7 & 32.5\\
    \hline
    \end{tabular}
    \caption{Relevant electrical parameters for the two types of HWRs and SSRs.}
    \label{tab:cav-params}
\end{table}
[floatfix]
\subsection{Mechanical Resonances}
In a rigid body treatment, SRF cavities have infinite sets of mechanical vibrational eigenmodes. These modes are described by their frequencies $\Omega_\mu =2\pi f_\mu$ and modal quality factors $Q_\mu$, where $\mu$ is the index number of a mechanical mode. Exciting the mechanical mode will cause the cavity to periodically deform, and thus modulate the cavity frequency. If any noise source in the cryomodule is periodic near $\Omega_\mu$, the mechanical mode can be excited and quickly shift the cavity frequency if $Q_\mu$ is sufficiently high i.e. the mode is poorly damped.

Cavity detuning can be broke down into two pieces: a quasi-static contribution ($\Dw_0$) and a time-dependent/dynamic contribution $\delta\omega$. Analogous to a string, every deformation of the cavity can be expressed as a sum of contributions from each relevant eigenmode. A result of this is that the total detuning $\Dw$ can be expressed as a sum of contributions from each mechanical resonance:
\begin{equation}
    \Dw=\Dw_0+\dw=\sum_\mu(\Dw_{0\mu}+\dw_\mu),
    \label{eq:det-sum}
\end{equation}
where $\Dw_{\mu0}$ and $\delta\omega_\mu$ are the modes static and dynamic contributions respectively. This treatment will allow us to break the cavity dynamics to more manageable pieces in the next section.

Medium-velocity, coaxial resonators present an interesting case when it comes to mechanical resonances. The effective lengths are still shorter compared to their higher-velocity counterparts, and the frequency is higher than low-velocity structures. This combination results in more compact, stiffer structures with higher mechanical resonance frequencies. All four of the resonators in \cref{tab:cav-params} have their first mechanical modes higher than 100~Hz. This is ideal from a microphonics standpoint as most cryomodule noise lies in the 0-100~Hz range, and presents an interesting angle to consider in terms of ponderomotive instability.

\subsection{Lorentz Force Detuning}
Due to the high electromagnetic fields in the cavity, a non-negligible radiation pressure $\vec{P}$ is applied to the cavity walls; with vector form as the following:
\begin{equation}
    \vec{P} = \frac{1}{4}(\mu_0 H^2-\epsilon_0 E^2)\hat{n},
\end{equation}
where $H$ and $E$ are the magnetic and electric fields respectively and $\hat{n}$ is the cavity surface normal unit vector. This pressure results in squeezing of the accelerating gap and outward expansion of the high magnetic field regions. In steady state cases, this pressure produces a frequency shift related to the square of the accelerating electric field. This is referred to as the static Lorentz force detuning (LFD) effect,
\begin{equation}
    \Dw_0 = -k_L E_{acc}^2,
    \label{eq:slfd}
\end{equation}
where $k_L$ is referred to as the static Lorentz force detuning coefficient. This factor is dependent on cavity geometry as well as stiffness \cite{sns-lfd, stiffness-lfd}. For continuous wave (CW) machines, this effect can be countered through the use of mechanical tuners on field ramping and can be eliminated for regular operation.

This Lorentz force detuning effect can also appear in dynamic cases, with each mechanical mode contributing to a portion of the detuning. Using an elastic model as well as Slater's theorem, one can arrive at the linearized equation of motion for the detuning contribution by mechanical mode $\delta\omega_\mu$ for deviations in the square of $E_{acc}$ \cite{schulze, delayen}:
\begin{equation}
    \ddot{\delta\omega_\mu} + \frac{\wmu}{Q_\mu}\dot{\delta\omega_\mu}+\Omega_\mu^2\delta\omega_\mu
    =-k_\mu\Omega_\mu^2\delta (E_{acc}^2).
    \label{eq:dlfd-td}
\end{equation}

In \cref{eq:dlfd-td}, $k_\mu$ is the modal Lorentz force detuning coefficient. This factor describes the overlap of the Lorentz pressure displacement with the mechanical mode shape.

Applying a Laplace transform on \cref{eq:dlfd-td}, one can solve for the transfer function for a single mechanical mode:
\begin{equation}
    G_\mu(s)=\frac{\delta\omega_\mu}{\delta(E_{acc}^2)}=\frac{\dw_\mu}{2E_{a0}\delta E_{acc}}=\frac{-k_\mu\Omega_\mu^2}{s^2+\frac{\wmu}{Q_\mu} s +\Omega_\mu^2}
    \label{eq:Gmu}
\end{equation}
where $E_{a0}$ is the steady-state accelerating electric field. The total electromechanical interactions of the cavity can be described through the Lorentz transfer function $G_L(s)$, which is the sum of all mechanical mode transfer functions (\cref{eq:Gmu}).
\begin{equation}
    G_L(s)=\sum_{\mu=1}^\infty G_\mu(s)
    \label{eq:LTF}
\end{equation}

\begin{figure}
    \centering
    \includegraphics[width=\linewidth]{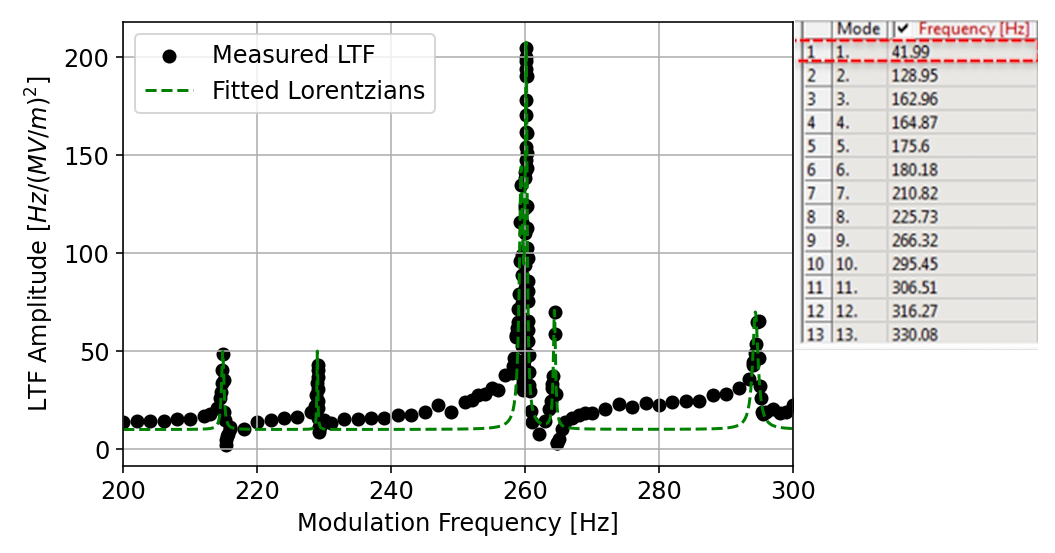}
    \caption{LTF amplitude v. modulation frequency for a FRIB $\beta=0.53$ HWR (left) and simulated mechanical modal report for the $\beta=0.53$ HWR (right). The circled frequency represents the fundamental mode for the cavity assembly.}
    \label{fig:b53-ltf}
\end{figure}

When, $s\rightarrow0$ (i.e. the static case), the gain of this electromechanical system becomes the static Lorentz force detuning coefficient:
\begin{equation}
    G_L(s=0)=\sum_\mu^\infty G_\mu(s=0)=\sum_\mu^\infty-k_\mu=-k_L.
\end{equation}
The Lorentz transfer function (LTF) can be measured in cavities by modulating the forward power and recording the field modulation and detuning amplitudes as the modulation frequency is swept. The Lorentz transfer function amplitude has been measured for a $\beta=0.53$ half-wave resonator at FRIB, and can be seen in \cref{fig:b53-ltf}. Active efforts are underway to capture LTF data for the PIP-II resonators at the Spoke Test Cryostat (STC) \cite{stc,stc2}. The mechanical mode quality factors $Q_\mu$ can be calculated from the $-3$~dB points of their respective peaks. With $Q_\mu$, $k_\mu$ can be calculated with the maxima of the Lorentz transfer function.

\subsection{Ponderomotive Instability}
In the presence of these electro-mechanical interactions, the cavity is opened up to unstable behavior via the Lorentz transfer function. Historically, this leads to two types of instabilities which we will now discuss.

\subsubsection{The Monotonic Instability}
\begin{figure}
    \centering
    \includegraphics[width=0.9\linewidth]{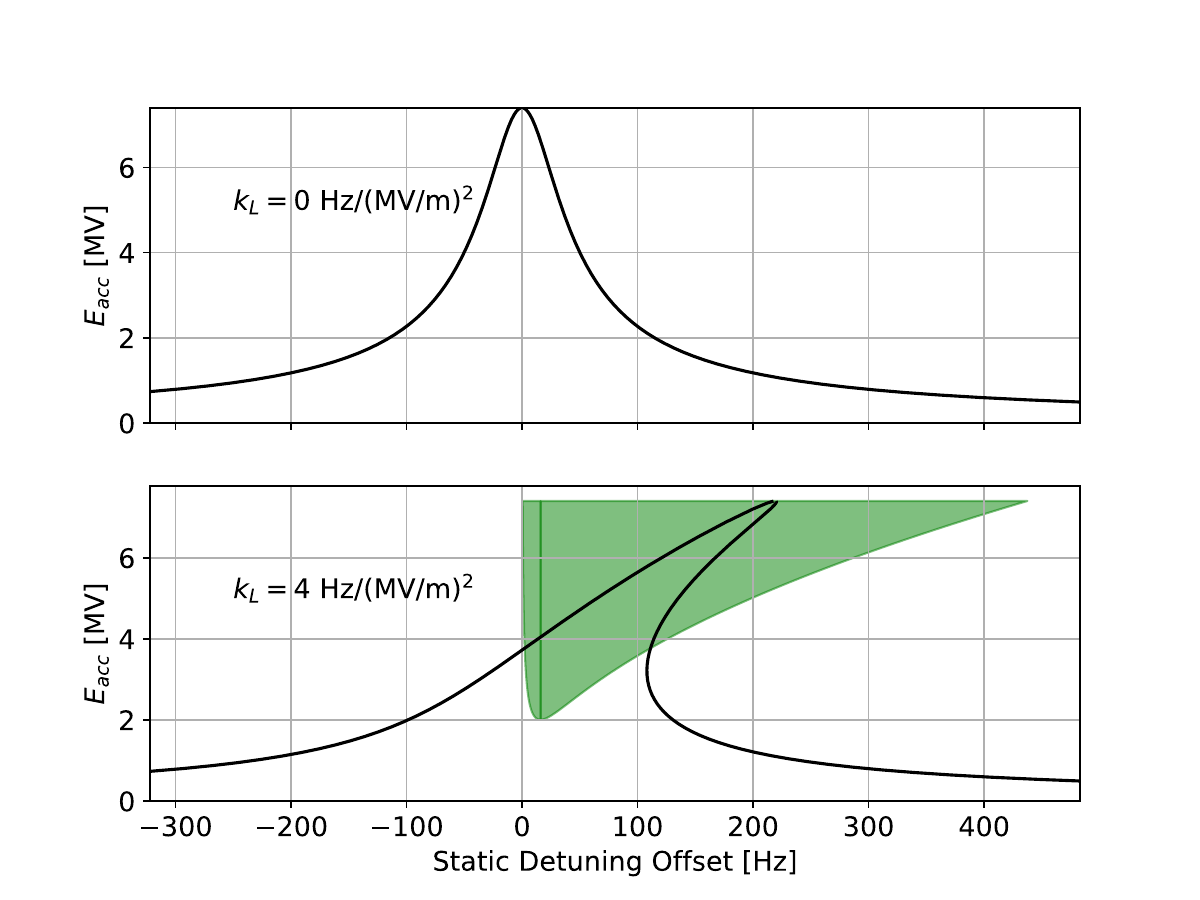}
    \caption{Calculated open-loop resonance curves for the two cases: with $k_L=0~\text{Hz/(MV/m)}^2$ (top) and with $k_L=4~\text{Hz/(MV/m)}^2$ (bottom). $Q_L=1\times10^7$ with $E_{max}=7.4$~MV/m in this example to showcase the tilting effect. Green shaded region represents unstable working points.}
    \label{fig:res-curves}
\end{figure}

The monotonic or jump instability is a direct result of the tilting of the resonance curve. The tilt creates a hysteresis region on the positive side of the resonance curve. An example calculation using parameters for a $\beta=0.53$ HWR can be seen in \cref{fig:res-curves}. Two cases are presented there: (1) the resonance curve for a cavity coupled at $Q_L=1\times10^7$ with no Lorentz force considered, and (2) the resonance curve with the Lorentz force considered. 

 This instability is characterized by a drop in field from a higher amplitude working point to the corresponding lower amplitude in the hysteresis region when perturbed. As such, this instability should only be encountered in the hysteresis region. For CW machines, the use of a mechanical tuner and tight cavity amplitude control can eliminate the risk of exciting this instability for all but the most extreme disturbances.

 For pulsed machines, the cavities undergo rapid, repeated detuning from the Lorentz force due to fast field fill-times and high repetition rate. For these cases, the monotonic instability poses the greatest risk to the cavity. Feed-forward control algorithms applied to a mechanical tuner as well as strict cavity amplitude control during the pulse flat-top time are used to minimize the risk of instability that comes with this mode of operation \cite{osti_1969329, Schappert:2011zz, pischalnikov-THPPR012, Schappert:IPAC2015-WEPTY036, PFEIFFER20201331}.

\subsubsection{The Oscillatory Instability}
 The oscillatory instability is a result of modulations of the cavity field, occurring in both open and closed loop. When the static detuning $\Dw_0$ passes the instability threshold, the loop transfer function switches from negative to positive feedback and the oscillations in field and detuning grow until an interlock trip or the oscillations are limited by some non-linear mechanism. We presented a more detailed schematic of the instability behavior in \cite{brown:hiat2025-wep19}.

\begin{figure}
    \centering
    \includegraphics[width=0.9\linewidth]{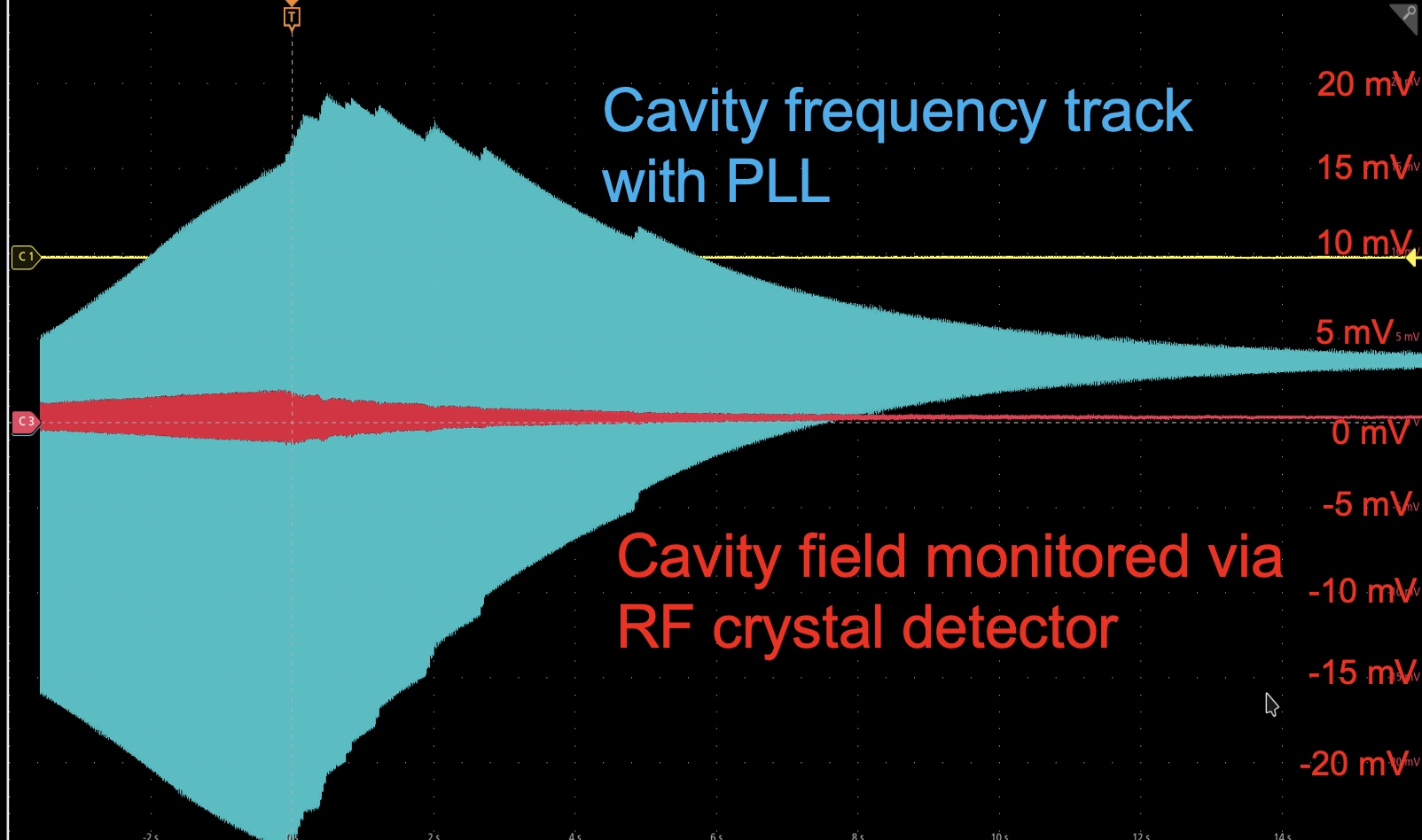}
    \caption{Oscilloscope screenshot of oscillatory instability onset in a $\beta=0.53$ HWR in the FRIB driver linac. The blue trace is a phase-lock-loop (PLL) error signal which shows cavity detuning. The red trace is the voltage from RF crystal detector on the cavity pick-up power.}
    \label{fig:pdi-epx}
\end{figure}

An experimental demonstration of this phenomena for a $\beta=0.53$ HWR in the FRIB driver linac can be seen in \cref{fig:pdi-epx}. When the cavity was detuned to the negative side of the resonance curve, there is a characteristic growth in field and detuning oscillations. When the cavity is tuned back to the positive side, the oscillations decay. In this case, the characteristic frequency excited is that of the strongest mechanical mode at $\sim$260~Hz. 

Amplitude low-level RF (LLRF) feedback systems can help to suppress this instability alongside phase control. For proportional/linear feedback systems, it is possible to push the instability threshold arbitrarily higher with larger control gains \cite{schulze}. 

 \subsection{Previous Work}
Examinations of the ponderomotive instabilities were first published by Karliner, Shapiro, and Shekhtman in 1967 for room temperature RF cavities \cite{karliner_instability_1967, karliner_vibration_1970}. Stability criteria were derived for room temperature cavities using the key assumption that $\Omega_\mu\tau\ll1$ for the relevant equations of motion. As we move to the superconducting regime, this approximation no longer holds, especially for stiffer geometries that have higher frequency mechanical modes. 

 The first general work for generator driven SRF resonators was produced by Schulze in 1971 \cite{schulze}. Using the transfer function method, general stability criteria were derived for the open loop system. Schulze also provided stability criteria for proportional feedback systems. Seven years later, Delayen would publish his studies on the stability of cavities driven in self-excited loop (SEL) operation \cite{delayen, DELAYEN20061}, showing that the open-loop SEL driven cavity is free of ponderomotive instabilities.

Work on state-space derivations for cavity stability has also been carried out at TRIUMF for generator driven systems. This included single cavity analysis as well as for vector-sum control systems \cite{Koscielniak:IPAC2019-THPRB010, Koscielniak:IPAC2019-THPRB009, Fong:IPAC2019-WEPRB003}. Additionally, analytical treatment has been done including frequency control systems for the European XFEL project \cite{BELLANDI2020361, bellandi-thesis}.

In 2025, we presented initial results of measurements of instability thresholds for the $\beta=0.53$ half-wave resonators and comparison with a cavity plus feedback simulation results using a Simulink model \cite{brown:hiat2025-wep19}. In this work we noted some behavior that deviates with established theory regarding the oscillatory instability. One deviation was the behavior of the instability threshold as the amplitude controller gain was varied: the instability threshold decreased at first then increased again as the controller gain was increased. It is predicted for proportional feedback systems that the instability threshold can be pushed further with increasing gain for generator driven cavities. 

Another deviation was in the presence of a second instability on the opposite side of the resonance curve. For cavities operating in CW mode with frequency and amplitude feedback control, we expect negligible risk of encountering the monotonic instability. Thus this instability is peculiar in its existence. 

Later in 2025, the RF team at IMP published their study into the oscillatory instability \cite{qiu-imp}. They present a exhaustive theoretical treatment of linearized cavity dynamics with and without the addition of linear feedback systems. We summarize the derived closed-loop characteristic polynomial (for a single mechanical mode):
\begin{equation}
\label{eq:char-poly}
\begin{split}
    [s+\wh(1+\cos{\psi_0}\cdot G_A(s))][s+\wh(1+G_\phi(s))]\\
    +\Dw_0(1+G_\phi(s))[\wh\sin{\psi_0}\cdot G_A(s)+\Dw_0]\\
    +2V_{a0}^2\Dw_0G_\mu(s)(1+G_\phi(s))=0,
\end{split}
\end{equation}
where $G_{A/\phi}(s)$ are the amplitude and phase controller transfer functions respectively, $V_{a0}=E_{a0}l_{eff}$ is the steady state cavity voltage, and $G_\mu(s)$ is \cref{eq:Gmu}. The real-parts to the roots of this equation determine the stability of the system: roots with positive real parts indicate instability. 
Their analytical model and simulation model showed great agreement with experimental onset of the oscillatory instability for proportional-integral controlled systems for heavily loaded ($Q_L\sim10^5-10^6$) HWRs at CAFE-2. 

This work finds its home in two places. First, we will showcase the presence of the oscillatory instability in phase-locked, self excited cavities at the STC. We will demonstrate preliminary stability analysis using a new method inspired by Karliner, Shapiro, and Shekhtman and discuss the avenues this opens for future studies. Second, we expand on the work in \cite{brown:hiat2025-wep19}: we showcase expanded instability threshold measurements and comparisons with our Simulink model. We use the analytical results of \cite{qiu-imp} and active disturbance rejection control transfer functions to explain the behavior of the instability with non-linear feedback systems. Finally, we will expand on the first analysis showed in \cite{qiu-imp} regarding amplitude and phase ($A/\phi$) v. $I/Q$ control and show how $A/\phi$ control can lead to lower oscillatory instability thresholds and/or controller instability.

\section{STC Studies}
\subsection{Experimental Observations}
Horizontal testing of individual cavities equipped with high power couplers is being carried out at the spoke test cryostat at Fermilab. Cavities are powered in this configuration by a self-excited loop with no amplitude control. A phase-lock-loop (PLL) is used to track the cavity frequency with respect to STC master clock and modulate the drive frequency.

Depending on the availability, sometimes cavities are tested without their piezoelectric tuner and instead equipped with mock tuner and/or safety brackets. While the tuner is not necessary for operation of the cavity in SEL mode, the reduced stiffness results in a larger LFD effect. On two occasions, we have observed ponderomotive oscillations in high-power testing of SSR2 cavities.

\begin{figure}
    \centering
    \includegraphics[width=\linewidth]{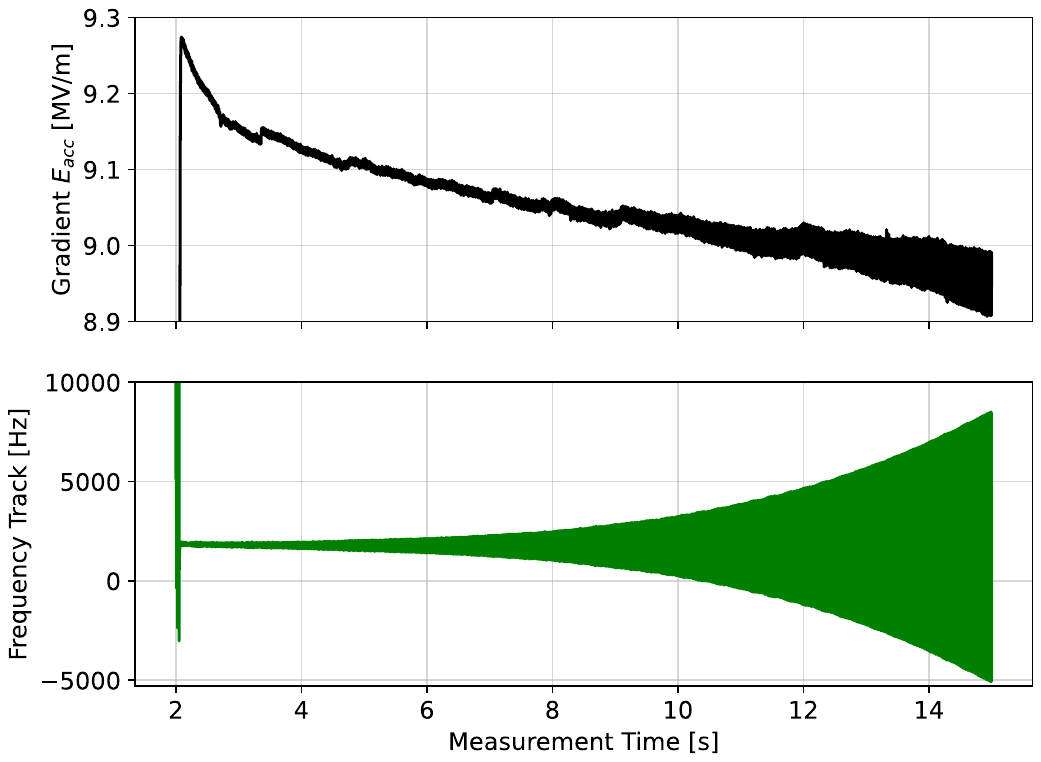}
    \caption{Cavity accelerating gradient $E_{acc}$ (top) and cavity frequency tracking in Hz (bottom) v. time at the instability threshold field for an SSR2 cavity at the STC.}
    \label{fig:ssr2-field}
\end{figure}

In both cases, the cavities had a measured static LFD coefficient of $k_L=7~\text{Hz/(MV/m)}^2$. In one case, the onset of ponderomotive oscillations occurred at $E_{acc}=9$~MV/m and in the other at $E_{acc}=$7~MV/m. An example of the oscillations in cavity field and frequency after passing the field threshold for an SSR2 cavity can be seen in \cref{fig:ssr2-field}. We can see that within $\sim12$~s the field oscillations have grown to $0.1$~MV/m peak-to-peak.

\begin{figure}
    \centering
    \includegraphics[width=\linewidth]{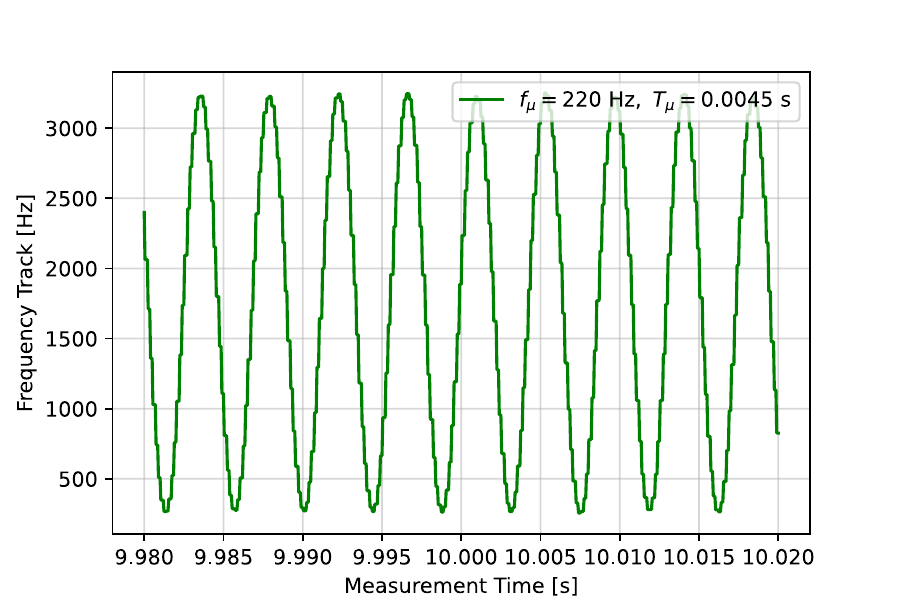}
    \caption{Cavity frequency-track value in Hz (offset from generator) v. measurement time during a test for an SSR2 cavity at the STC.}
    \label{fig:detuning-fwf}
\end{figure}

In this case this lead to 13~kHz peak-to-peak detuning, again showing how strong these oscillations can be. On a separate measurement, we were able to determine the frequency of the oscillations. A slice of the cavity frequency-track waveform can be seen in \cref{fig:detuning-fwf}. We see that a 220~Hz mechanical mode is excited, a frequency well above the usual microphonics range. 

The threshold of these oscillations also changed when the gain of the phase shifter was changed by 2~dB. This again tells us that these are in-fact ponderomotive oscillations, which are sensitive to phase control systems. We have not observed these oscillations in the SSR1 variety. SEL mode is usually considered better in terms of preventing ponderomotive instability due to better amplitude stability compared to generator driven systems. These factors combined with the relatively high frequency of the instability driving mode has led us to consider which design and mechanical parameters most strongly affect the stability of the cavity. We outline our initial strategy in the following section.

\subsection{Mathematical Modeling}
 We have seen that even in phase-locked SEL cavities, these oscillations can still appear. Following our observations, we are now pursuing a detailed analytical model to examine the oscillatory instability from cavity design perspective as means to better understand which factors are most critical in determining stability. We outline our initial progress in this section. Let us start from the envelope equations for deviations in the in-phase ($I$) and quadrature ($Q$) components and separating the detuning into the static and dynamic pieces in open-loop. Neglecting constant generator terms, we have:
 \begin{align}
    \dot{\dI}+\wh\dI+\Dw_0\dQ=0,\label{eq:dot-di}\\
    \dot{\dQ}+\wh\dQ-\Dw_0\dI+V_0\dw=0\label{eq:dot-dq}.
 \end{align}

We assume that the on-resonance, steady-state cavity voltage is in-phase with the generator,
that is $V_0=I_0$. If we take a single mechanical mode treatment to start, then \cref{eq:dlfd-td} becomes the only equation for the detuning. Let us have state vector $\mathbf{x}$ defined as,
\begin{equation}
    \mathbf{x}=[\dI,~\dQ,~\dw_\mu,~\dot{\dw_\mu}]^T,
\end{equation}
then using \cref{eq:dot-di}, \cref{eq:dot-dq}, and \cref{eq:dlfd-td} we can construct the state matrix of the un-driven system:
\begin{equation}
    A=
    \begin{bmatrix}
        -\wh& -\Dw_0& 0&0\\
        \Dw_0 & -\wh & V_0 & 0\\
        0 & 0 & 0 & 1\\
        -2k_\mu\wmu^2V_0 & 0 & -\wmu^2 & -\frac{\wmu}{Q_\mu}
        \end{bmatrix}.
        \label{eq:state-mat}
\end{equation}

The stability of the system can then be determined by the eigenvalues of \cref{eq:state-mat}:
\begin{equation}
    \det{(I-sA)}=0,
\end{equation}
where an eigenvalue with a positive real portion denotes unstable behavior. What results is a 4th order polynomial with coefficients:
\begin{align}
    a_4=1,\\
    a_3=\frac{\wmu}{Q_\mu}+2\wh,\\
    a_2=\wmu^2+\wh^2+\Delta\omega_0^2+\frac{2\wh\wmu}{Q_\mu},\\
    a_1=2\wh\wmu^2+\frac{\wmu}{Q_\mu}\left(\wh^2+\Delta\omega_0^2\right),\\
    a_0=\wmu^2(\Delta\omega_0^2+\wh^2-2\Delta\omega_0k_\mu V_0^2),
\end{align}
where $a_i$ is the coefficient for the i'th order of $s$. These coefficients also fall out from the derivation in \cite{qiu-imp} and \cite{schulze}. The $a_0$ coefficient relates to the monotonic instability (the $s=0$ case). The thresholds for the monotonic instability are then the conditions in which $a_0=0$. This gives a relatively straightforward criterion through use of the quadratic formula:
\begin{equation}
    \Dw_{\pm}=k_\mu V_0^2\pm\sqrt{k_\mu^2V_0^4-\wh^4}.
    \label{eq:mono-thresh}
\end{equation}
 Static detuning offset within $\Dw_-<\Dw_0<\Dw_+$ correspond to unstable working points for $V_0\geq\sqrt{\wh/k_L}$. In terms of the oscillatory instability, there are analytical forms for thes stability criterion for single mode systems as demonstrated by Schulze \cite{schulze}. Numerical solutions can be found for given mechanical factors with estimation of the roots or through Routh-Hurwitz analysis to find the thresholds for the oscillatory instability in multi-mode systems. 

One path to an analytical stability criterion is to then examine the extrema of the system, of which can be found through the roots of the derivative:
\begin{equation}
    4a_4s^3+3a_3a^2+2a_2s+a_1=0.
    \label{eq:cubic}
\end{equation}
Cardano's method can then be used to find the roots of this expression. This method relies on four key expressions; the first two relate to depressing the cubic:
\begin{align}
   Q=\frac{8a_4a_2-3a_3^2}{48a_4^2}, \\
   R=\frac{4a_4a_3a_2-8a_4^2a_1-a_3^3}{64a_4^3},
\end{align}
and the final components of the cubic roots:
\begin{align}
    S=\sqrt[3]{R+\sqrt{R^2+Q^3}},\\
    T=\sqrt[3]{R-\sqrt{R^2+Q^3}}.
\end{align}

The real portion of the roots to \cref{eq:cubic} are then given by $\pm(S+T)$. While these expressions are complex, we believe that the extra terms will give us more space to examine how each parameter affects the stability of the cavity. Schulze has derived an analytical criterion for the oscillatory instability following Routh-Hurwitz analysis, and we can use that result to verify our analysis in single mode cases.

\section{Closed Loop Studies}
\subsection{Active Disturbance Rejection Control}
Constant cavity amplitude and phase is essential for the successful acceleration of charged particles. As such, all accelerators will make use of feedback systems in their LLRF systems to achieve this criterion. The proportional-integral (PI) controller algorithm is popular due to the wide availability of its technology and its simplicity. Accelerator laboratories such as Fermi National Accelerator Lab, the Stanford Linear Accelerator Complex, the Spallation Neutron Source, and many more \cite{varghese_llrf_2023, doolittle-lcls2, sns-llrf, ma_virtual_2024, bouly_superconducting_2022}, utilize PI control algorithms for amplitude and phase stabilization. 

\begin{figure*}
    \centering
    \includegraphics[width=0.7\textwidth]{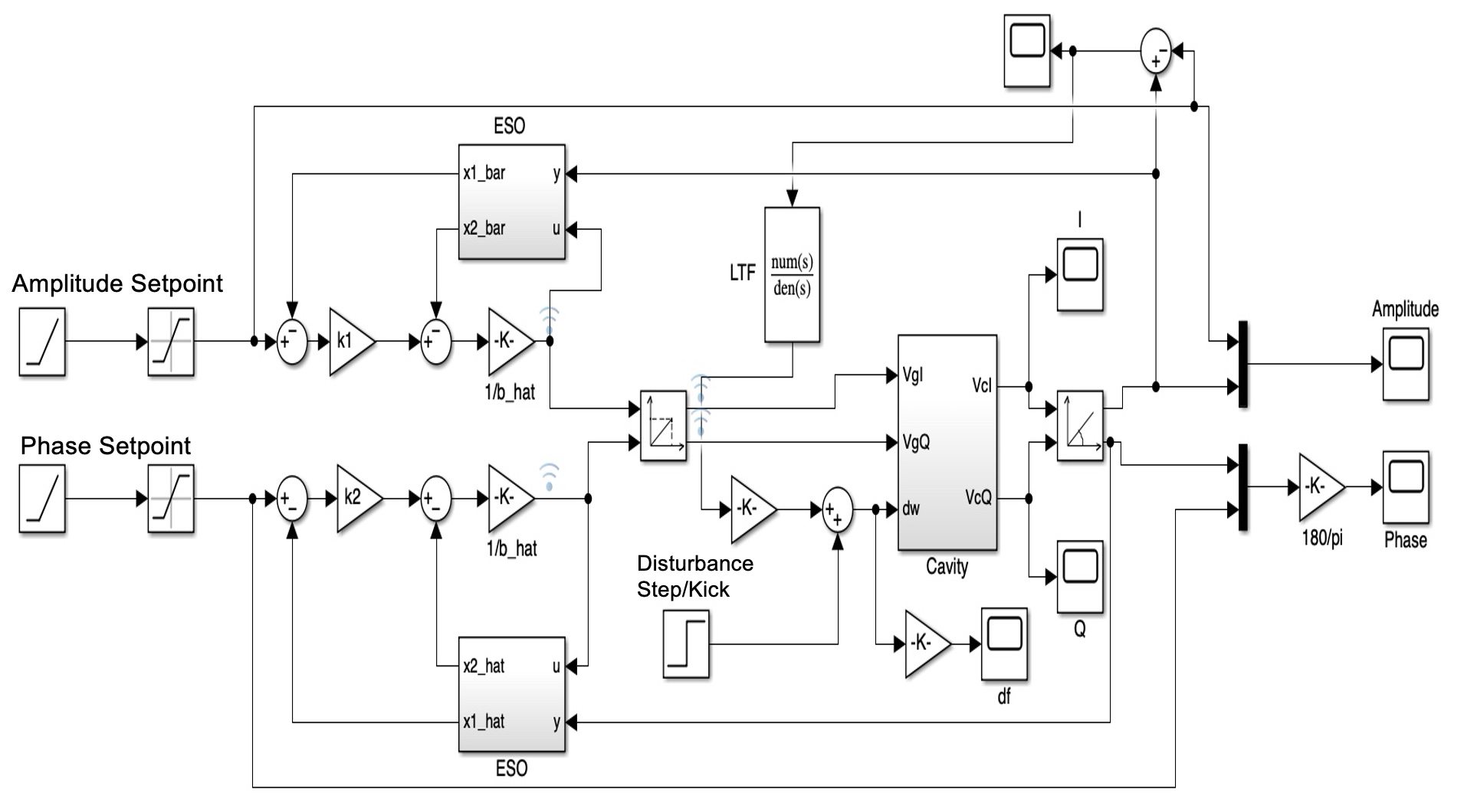}
    \caption{Schematic of Simulink model used to simulate the onset of the oscillatory instability threshold.}
    \label{fig:sim-schematic}
\end{figure*}

FRIB utilizes an active disturbance rejection control (ADRC) algorithm for amplitude and phase control \cite{zhao-fp}. ADRC allows for the treatment of internal plant dynamics and external disturbances as one generalized disturbance $f$ \cite{han-trans}. For example, a simple first order system can represented in the following time domain equations:
\begin{align}
    \dot{x}=f(x, d, t) + bu,
    \label{eq:simp-fo}\\
    y=x
\end{align}
where $x$ is the internal state, $d$ is the external disturbance, $u$ is the system input, and $b$ is the input coefficient. For a closed loop system with output $y$, we can construct a state-space model with state vector $\textbf{x}=[x, f]^T$. Then we can express the total dynamics of the system in the following matrix-vector equations:
\begin{align}
    \dot{\textbf{x}}=[\dot{x}, \dot{f}]^T=A\textbf{x}+B\textbf{u}+E\mathbf{\dot{f}}\\
    y=C\textbf{x}+D\textbf{u}
\end{align}
where, $A,~B,~C,~D$, and $E$ are the state, input, output, feed-through, and disturbance matrices respectively.

ADRC utilizes an extended state observer (ESO) to provide estimates of the state vector and output $\mathbf{\hat{x}}$ and $\mathbf{\hat{y}}$ respectively. The state-space equations for the observer follow closely to those of the dynamic systems:
\begin{align}
    \dot{\hat{\textbf{x}}}=A\hat{\textbf{x}}+B\textbf{u}+\mathbf{L}\cdot(\textbf{y}-\mathbf{\hat{y}}) \\ 
    \mathbf{\hat{y}}=C\mathbf{\hat{x}}+D\mathbf{u},
\end{align}
where $\mathbf{L}$ is the gain vector for the observer, which is a function of the observer bandwidth $\omega_{ob}$:
\begin{equation}
    \mathbf{L}=[3\omega_{ob},~3\omega_{ob}^2,~\omega_{ob}^3]^T.
\end{equation}
The formulation of the extended state observer changes a bit when we move to a discretized format. The exact formulation for the discrete $\mathbf{L}$ and extended state observer used in this work can be found \cite{disc-ESO}.

For the first order formalism, the input to the plant can then be formulated from the ESO's estimation of the state and disturbance as such:
\begin{equation}
    u = \frac{\omega_c(r-\hat{x}_1) - \hat{x}_2}{b_0},
    \label{eq:input}
\end{equation}
where $\omega_c$ is the controller bandwidth, $b_0$ is an approximation of the input coefficient $b$, $r$ is the reference signal, and $\hat{x}_{1,2}$ are the state and disturbance estimations respectively. In this formalism, it is clear the controller bandwidth serves as a proportional gain factor. 

There also exists transfer function representation for the linearized ADRC formalism. The forms for first and second order ADRC have been derived in \cite{zhao-nmp, tian-gao-tf}. We summarize the first order transfer function forms for the pre-loop filter,
\begin{equation}
    H(s) = \frac{\omega_c(s^2+\beta_1s + \beta_2)}{(\omega_c\beta_1 + \beta_2)s + \omega_c\beta_2},
    \label{eq:prefilter}
\end{equation}
as well as the controller:
\begin{equation}
    G_c(s)=\frac{\omega_c}{bs}\frac{(\beta_1+\frac{\beta_2}{\omega_c})s +\beta_2}{s + (\beta_1+1)},
    \label{eq:controller}
\end{equation}
where $\beta_1 = 2\omega_{ob}$ and $\beta_2=\omega_{ob}^2$. For stability analysis, $H(s)$ does not need to be considered as it is designed to be stable and is not part of the closed-loop, but we include the transfer function for completeness.

 For control of cavity amplitude and phase (or $I/Q$), the input coefficient estimation $b_0$ is the electrical half-bandwidth $\wh$. As we are using a first order system in the form of the envelope equations, we are using a first order ADRC control scheme as was originally developed for the Re-accelerator (ReA3) at FRIB (then the National Superconducting Cyclotron Laboratory)  \cite{VINCENT201111}.

ADRC and state observers have slowly worked their way into some accelerator controls. Disturbance observers have seen implementation in the LLRF controllers at KEK \cite{qiu-dob}. Similarly, a modified ADRC algorithm was applied to control a mechanical tuner for TESLA style cavities at ISOLDE at CERN \cite{adrc-tuner}.

\subsection{Simulink Simulation}
We are using a Simulink time domain model that was originally described in \cite{VINCENT201111}. MATLAB/Simulink allows for the implementation of continuous and discrete transfer functions. In \cite{brown:hiat2025-wep19}, we added the LTF to the model to simulate the electromechanical behavior of the cavity. A schematic of this model can be seen in \cref{fig:sim-schematic}. These simulations are ran with a loop rate of 200~kHz.

Since our previous study, we have changed the way we scan for the instability threshold. Previously, we would gradually change the detuning offset to different levels until we noted the appearance of oscillations within the simulation run time. While we simulate the system over a minute, this might not be a sufficient amount of time for exploring weakly coupled mechanical modes and/or lower cavity amplitudes. 

In real cavities within the linac, microphonics can help promote the instability once the threshold is crossed due to residual vibrations. In our Simulink model, we do not have residual vibrations. To ensure that we encounter the instability within the simulated time frame, we switched the detuning offset ramp to a sharp kick/step. This excites all the mechanical modes as if the cavity was struck, and the instability driving mode will either grow or decay depending on if the step level has passed the instability threshold.

\subsection{Experimental Measurements}
Measurements of the instability thresholds in terms of a static detuning offset can be done on half-wave resonators in the linac through the LLRF controller. A forward-cavity offset is used to denote the phase difference between the forward power and cavity when the cavity is tuned on resonance. By adding an adjustment to this offset, we can detune the cavity by introducing a detuning phase/angle $\psi_0$, which can be translated to rad/s as:
\begin{equation}
    \Dw_0=\wh\tan{\psi_0}.
\end{equation}

We can therefore add increments of detuning angle until we see a growth in oscillations measured cavity field, phase, and detuning that comes with the oscillatory instability. We recorded the thresholds as we varied the controller bandwidth $\omega_c$ for the amplitude control loop. The phase control gain parameters were left unchanged as the amplitude controller is the main prevention mechanism for the oscillatory instability in closed loop $A/\phi$ configuration. 

\begin{figure}
    \centering
    \includegraphics[width=\linewidth]{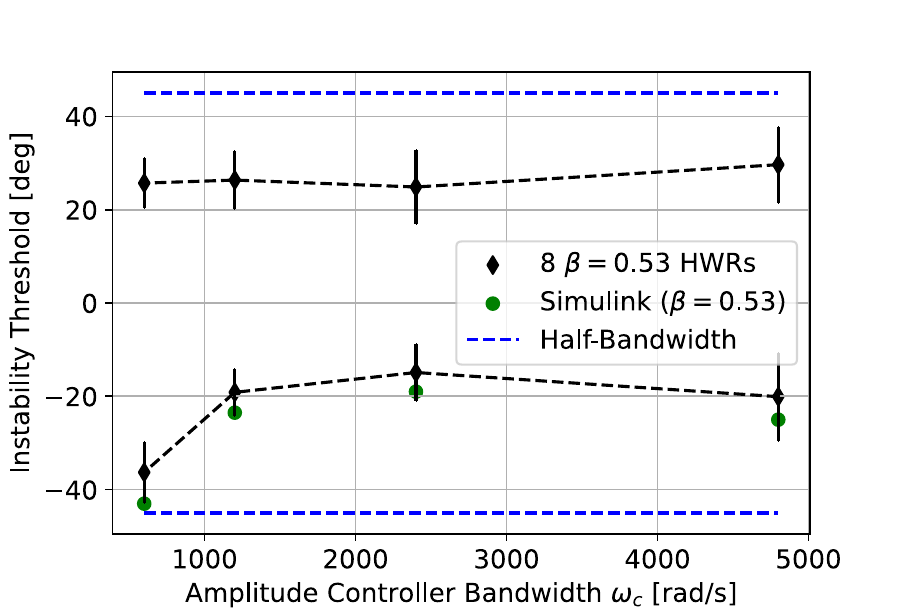}
    \caption{Instability thresholds for $\beta=0.53$ HWR in terms of detuning angle $\psi_0$ versus the amplitude controller bandwidth $\omega_c$. Measurements taken across eight HWRs within the FRIB driver linac.}
    \label{fig:b53-thresh-init}
\end{figure}

Similarly we can apply a detuning offset to the cavity in our Simulink model and record the onset of the same growth in detuning, field and phase. We then compare simulated and experimental results for the $\beta=0.53$ HWR in \cref{fig:b53-thresh-init}.

\begin{figure}
    \centering
    \includegraphics[width=\linewidth]{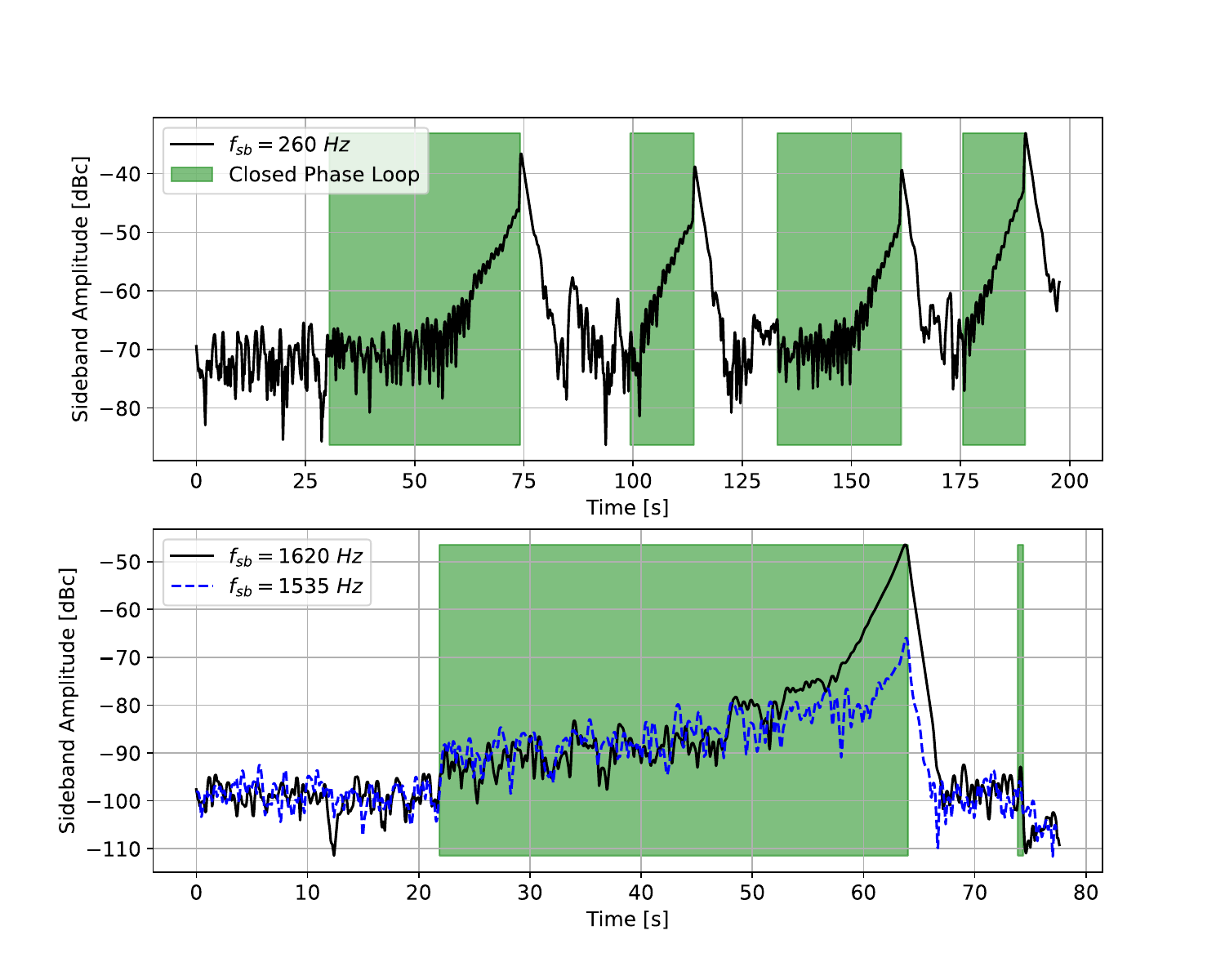}
    \caption{Plot of relevant instability driving sidebands from spectrogram data for negative detuning (top) and positive detuning (bottom) in a $\beta=0.53$ HWR within the FRIB driver linac. Green shaded regions represents time periods where the phase control loop is closed. Unshaded regions are phase-open regimes.}
    \label{fig:sbs}
\end{figure}

Some of the instability threshold measurements were accompanied with measurements of the low-frequency spectrum of the carrier during the instability event. Spectrograms were taken when the cavity was detuned positively and negatively. In both cases, sidebands denoting the growth of the instability driving frequency (frequencies) can be seen. On a few attempts, we were able to open and then re-close the phase control loop before an instability induced trip. It can be seen that the sidebands decay upon opening the phase control loop, confirming that this is not a strong microphonics event as microphonics would persist upon opening the phase loop. The relevant side band data can be seen in \cref{fig:sbs}. 

\begin{figure}
    \centering
    \includegraphics[width=\linewidth]{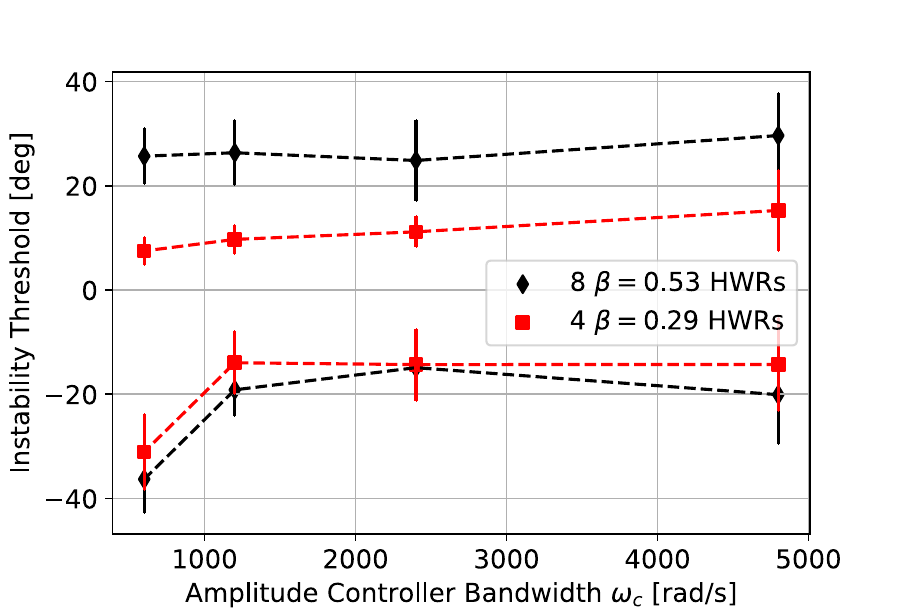}
    \caption{Instability threshold in terms of detuning angle v. amplitude controller bandwidth for 4 $\beta=0.29$ HWRs within the FRIB driver linac. Included are the measurements for the $\beta=0.53$ HWRs for comparison.}
    \label{fig:b29-init}
\end{figure}

We note that the trends seen in \cref{fig:b53-thresh-init} are not limited to only the $\beta=0.53$ half-wave resonators. We see similar behavior in the oscillatory instability threshold for the $\beta=0.29$ HWRs as well, which can be seen in \cref{fig:b29-init}. This indicates that this phenomena is not necessarily specific to one type of cavity only. Additionally, the Simulink model replicates this behavior which tells us that this is not an implementation effect.

\subsection{Analysis}
These experimental measurements of the $\beta=0.53$ HWR give us some insight into not only the accuracy of our model, but into the cavity dynamics as well. At the present, we do not have a viable LTF measurement for the $\beta=0.29$ HWR so our analysis and simulations will stay relegated to the $\beta=0.53$ variety. It is clear from \cref{fig:b53-thresh-init} that our Simulink model does well in predicting the instability threshold on average for negative detuning ($\psi_0<0$).

Additionally, we see that our Simulink model predicts no instability up to $45^\circ=\wh$ detuning, whereas an instability presents itself at an almost constant detuning angle across all controller bandwidths. In the presence of amplitude control, we expect that the risk of the monotonic threshold to be eliminated. We also see in \cref{fig:sbs} that\ this instability is accompanied with high-frequency sidebands at 1620~Hz and 1535~Hz. This high-frequency component is suspect as the filter-like nature of the cavity naturally attenuates high-frequency components internally, suggesting a external driver of an electrical nature.

\subsubsection{Analytical Modeling}
In order to better understand the phenomena we see in our measurements and simulation, we now carry out analytical modeling of the closed loop system. We will use a single-mechanical-mode treatment for ease of calculation and because we only see one instability driving mode during the measurements. We will also use the amplitude/phase form of the envelope equation as those are the FRIB control process variables. First, we will remark on the cross-talk between process variables. Treating the RF envelope equation with deviations from steady state, one will find the following two equations:
\begin{equation}
\begin{aligned}
    \dot{\dV}+\wh\dV+\Delta\omega_0V_0\dP&=\\-\wh(G_A*\dV)\cos{\psi}-\Delta\omega_0V_0(G_\phi*\dP)\label{eq:env-voltage}\\
    %\dot{\dP}+\wh\dP-\frac{\Delta\omega_0}{V_0}\dV-\dw&=\\\frac{-\wh}{V_0}\sin{\psi}(G_A*\dV)-\wh(G_\phi*\dP),
\end{aligned}
\end{equation}
\begin{equation}
\begin{aligned}
    %\dot{\dV}+\wh\dV+\Delta\omega_0V_0\dP&=\\-\wh(G_A*\dV)\cos{\psi}-\Delta\omega_0V_0(G_\phi*\dP)\label{eq:env-voltage}\\
    \dot{\dP}+\wh\dP-\frac{\Delta\omega_0}{V_0}\dV-\dw&=\\\frac{-\wh}{V_0}\sin{\psi}(G_A*\dV)-\wh(G_\phi*\dP),
\end{aligned}
\end{equation}
where $*$ represents the convolution operator. After applying a Laplace, we can solve for the transfer function that describes changes in voltage $\dV$ from changes in phase $\dP$ using \cref{eq:env-voltage}:
\begin{equation}
    \frac{\dV}{\dP}=\frac{-\Dw_0V_0(1+G_\phi(s))}{s + \wh(1+\cos{(\psi_0)G_A(s))}}.
    \label{eq:sens-tf}
\end{equation}

\begin{figure}
    \centering
    \includegraphics[width=0.95\linewidth]{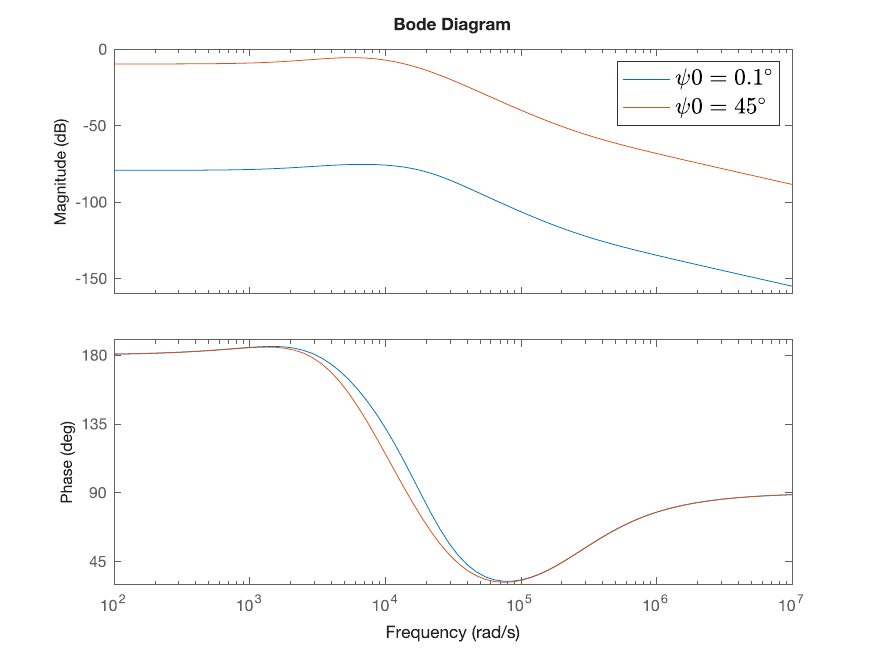}
    \caption{Bode plot for \cref{eq:sens-tf} for $\psi_0=0.1^\circ$ and $\psi_0=45^\circ$ respectively. The magnitude response for $\psi_0=0.1^\circ$ peaks at $f=1166~Hz$ and $\psi_0=45^\circ$ peaks at $f=856~Hz$.}
    \label{fig:dvdp-bode}
\end{figure}

As can be seen, the amplitude and phase controller information is now encoded in the cross-talk channel between the two variables. Like the open-loop system, this gain factor increases as the detuning increases, we can get the Bode plot of \cref{eq:sens-tf} using \cref{eq:controller} for the amplitude and phase controllers to see the frequency response of this crosstalk. The Bode plots for \cref{eq:sens-tf} for two cases can be seen in \cref{fig:dvdp-bode}. It is clear that adding the ADRC transfer functions adds a peak frequency which is more easily passed between $\dP$ and $\dV$. As the magnitude of the detuning increases, this peak shifts down in frequency.

\begin{figure}
    \centering
    \includegraphics[width=0.9\linewidth]{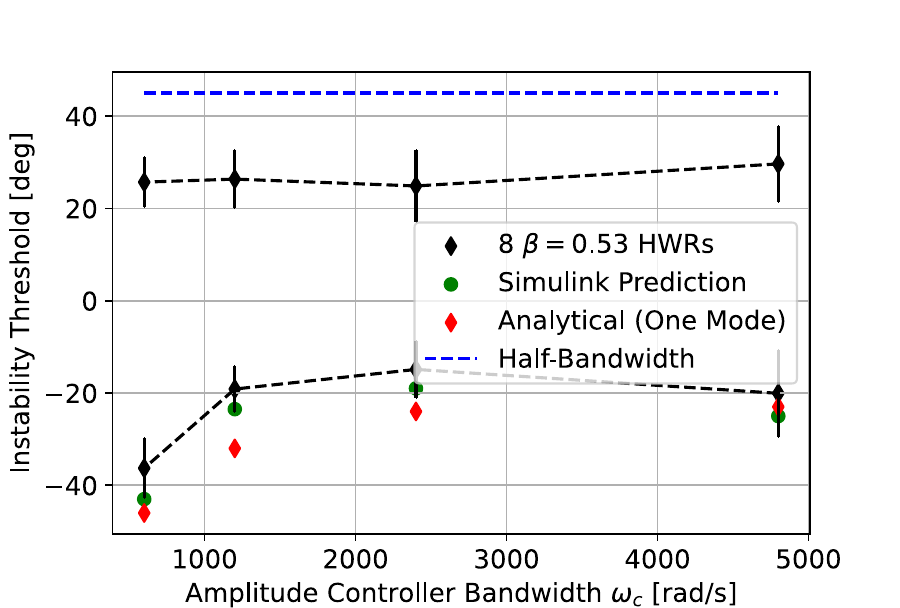}
    \caption{Plot of instability thresholds in terms of detuning angle v. the amplitude controller bandwidth $\omega_c$.}
    \label{fig:analytical-prelim}
\end{figure}

Using the results of \cite{qiu-imp}, we can find the characteristic polynomial for a single mechanical mode that includes the ADRC controllers (\cref{eq:controller}) and mechanical mode transfer function (\cref{eq:Gmu}). We list the relevant controller and mechanical parameters from the LTF in \cref{tab:anal-params}. We can then vary $\psi_0$ until one root of \cref{eq:char-poly} has a positive, real portion. We present the results of this analysis in \cref{fig:analytical-prelim}. 

\begin{table}[]
    \centering
    \begin{tabular}{c|c}
    \hline
        Parameter [units] & Value \\
    \hline
        $\omega_{c}$ [rad/s] & 600,1200,2400,4800 \\
         $\omega_{ob,A}$ [rad/s] & 19200 \\
         $\omega_{c,\phi}$ [rad/s] & 600 \\
         $\omega_{ob,\phi}$ [rad/s] & 9600 \\
         $f_\mu$ [Hz] & 259.3 \\
         $Q_\mu$   &   400 \\
         $k_\mu$ [$Hz/(MV/m)^2$] & 0.33\\
    \hline
    \end{tabular}
    \caption{List of controller and observer bandwidths used for the $A/\phi$ controllers as well as the relevant cavity mechanical parameters for the instability driving mode.}
    \label{tab:anal-params}
\end{table}

\begin{figure}
    \centering
    \includegraphics[width=0.95\linewidth]{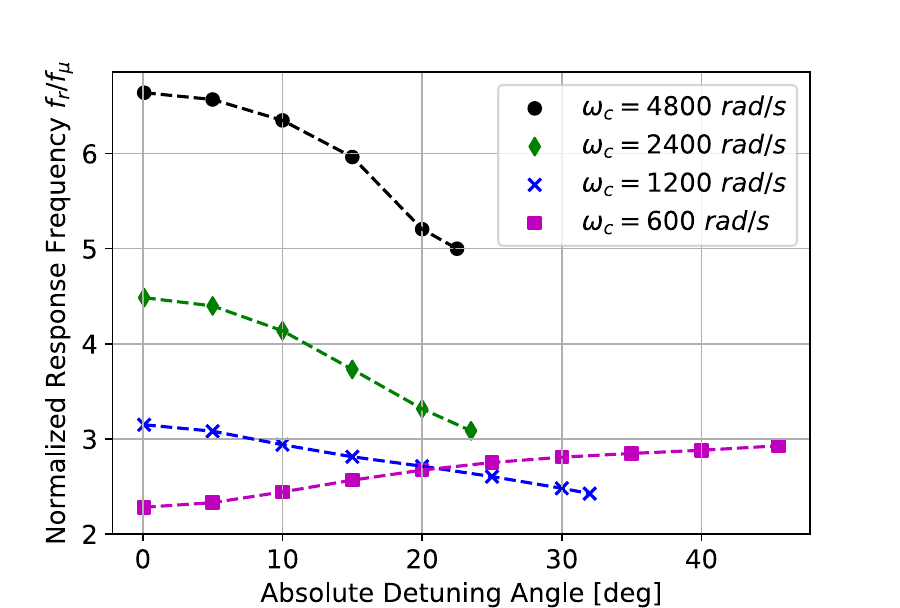}
    \caption{Plot of the combined controller frequency response $f_r$ normalized to the instability driving mechanical mode $f_\mu$ v. the absolute value of the detuning angle.}
    \label{fig:fres-fmu}
\end{figure}

We find good agreement with the analytical model between both the measurements and Simulink predictions for $\psi_0<0$, and again we see that the instability threshold decreases with increasing $\omega_c$. This suggests that this effect is a result of coupling between the ADRC and the electromechanical system. The other roots of \cref{eq:char-poly} detail the frequency response of the system. As stated, the ADRC controllers add a characteristic response frequency $f_r$ to the system's total response. If we examine the changes in $f_r$ as we detune the cavity up to the instability threshold we see how this effect manifests. If the controller frequency response approaches a low enough harmonic of the mechanical mode $f_\mu$, higher harmonics are excited and lead to earlier instability for higher $\omega_c$. We demonstrate this relationship in \cref{fig:fres-fmu}.
[floatfix]

\begin{figure}
    \centering
    \includegraphics[width=0.95\linewidth]{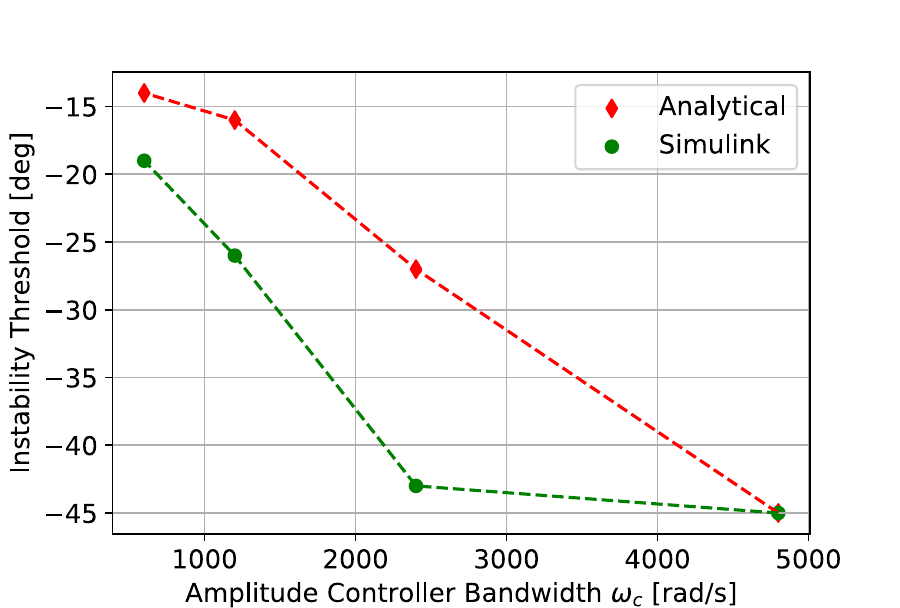}
    \caption{Predicted instability threshold v. amplitude controller bandwidth when the instability driving modes have frequencies shifted down to 90~Hz.}
    \label{fig:fshift}
\end{figure}

We note that this is not an effect purely on the use of ADRC as a feedback controller. This effect of reducing the instability threshold as an increase in controller gain is a by-product of the ADRC \textbf{and} the high frequency of the instability driving mechanical mode. If we use the same mechanical parameters as in \cref{tab:anal-params} but instead we drop the instability mechanical resonance frequency to 90~Hz, both our analytical and Simulink model show the expected behavior (\cref{fig:fshift}).

\begin{figure}
    \centering
    \includegraphics[width=0.95\linewidth]{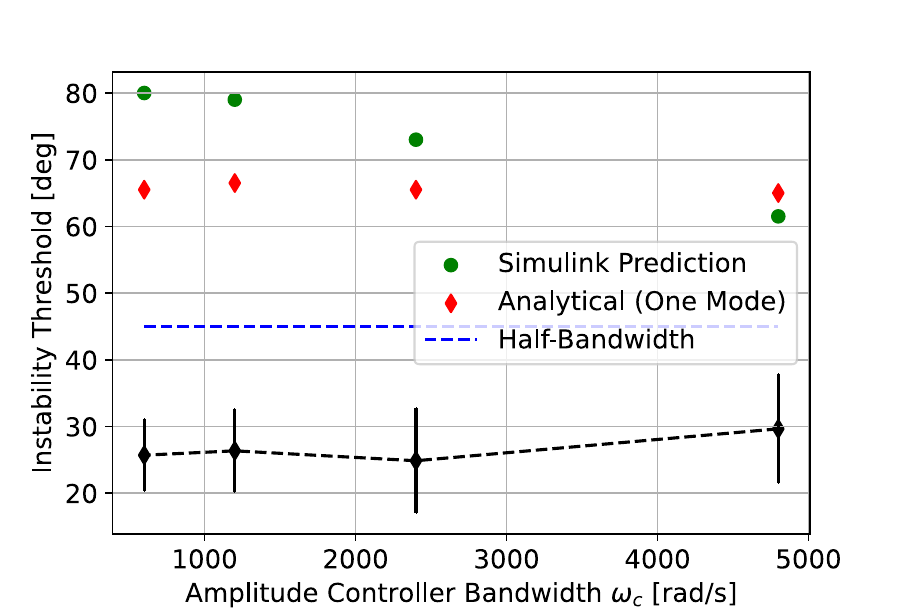}
    \caption{Instability thresholds: measured averages, Simulink, and analytical v. the amplitude controller bandwidth for the positive side instabilities.}
    \label{fig:pos-side-comp}
\end{figure}

If we turn our attention to the other side of the resonance curve, one question remains: why do our models not predict the instability we see on real $\beta=0.53$ HWRs on the positive side of the resonance curve? If we extend our region of search to beyond the half-bandwidth region, we see that our model does show instability thresholds, but to a much larger detuning than we see in the real machine. The results of this comparison can be seen in \cref{fig:pos-side-comp}.

The analytical model shows that the instabilities on this side of the resonance curve are driven with high-frequency signals ($>1$~kHz). We see this phenomena in the sideband data from the online measurements, suggesting that these predicted and real instabilities stem from the same cause. The high-frequency signals suggest that this instability is related to the ADRC controllers themselves and electrically/digitally driven rather than electro-mechanically driven. These instability thresholds do not shift when we shift the instability driving mode to 90~Hz, furthering that these instabilities are related purely to the controller and not the electromechanical system.

\subsection{Comparisons}
\subsubsection{Comparison with I/Q Control}
FRIB LLRF control involves the control of amplitude and phase directly. Another method to carry out this control is by controlling the in-phase and quadrature components of the cavity voltage phasor:
\begin{equation}
    \tilde{V_c}=V_ce^{j\phi}=I+jQ.
    \label{eq:iq-ap}
\end{equation}
Both methods provide pros and cons, and the choice of control variables must be made on a machine base level. $I/Q$ control can cover all four quadrants of the complex plane, but $A/\phi$ control allows for more precise control of these values. It is important to remark that the phase is insensitive to changes in amplitude, but the vice-versa is not necessarily true. It is also worth mentioning that in $A/\phi$ control, large disturbances can result in the phase being driven to the next quadrant. In order to counter this, phase change limiters are utilized.

We seek to compare the two methods of control in terms of preventing the onset of the oscillatory instability. In \cite{qiu-imp}, the authors present a preliminary analytical comparison of instability thresholds under $A/\phi$ and $I/Q$ control. The authors show that under $A/\phi$ control, the amplitude controller response terms are damped by $\cos\psi_0$ and $\sin\psi_0$, reducing their effectiveness. Additionally, the phase controller is coupled to the electromechanical system (\cref{eq:char-poly}).

\begin{figure}
    \centering
    \includegraphics[width=0.95\linewidth]{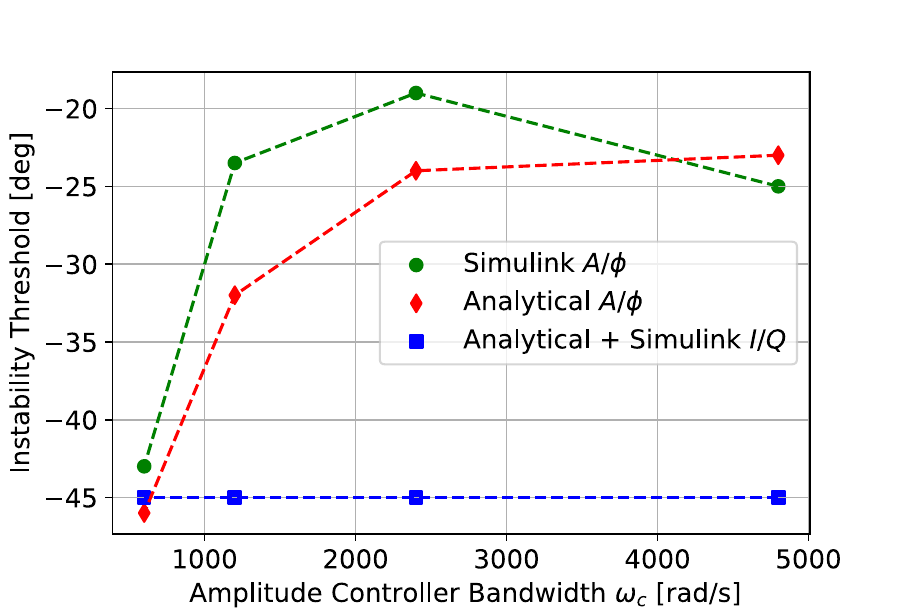}
    \caption{Instability threshold v. amplitude controller bandwidth comparing $A/\phi$ and $I/Q$ controllers.}
    \label{fig:iq-comp}
\end{figure}

They demonstrate that in $I/Q$ control these damping and cross terms are eliminated, which resulted in higher instability thresholds. We carry out this analysis and show this is true even with linear and non-linear ADRC controllers with the results in \cref{fig:iq-comp} where we look for instabilities up to the negative half-bandwidth. We report that the $I/Q$ controlled systems are oscillatory instability free (even up to $\psi_0=-89^\circ$ where the detuning approaches infinity). Additionally, the positive side is instability free in $I/Q$ control up to $\psi_0=89^\circ$ as well. 

The removal of the instability from the positive side tells us that this controller instability lies in the choice of control variables. We now expand on the findings of \cite{qiu-imp} to explain how these damping and cross-coupling terms affects the response of the system in the presence of disturbances and how this can lead to instability. As stated earlier, deviations in phase can translate to deviations in amplitude via \cref{eq:sens-tf}. An analogous transfer function can be derived for the $I/Q$ formalism:
\begin{equation}
    \frac{\delta I}{\delta Q}=\frac{-\Dw_0}{s+\wh(1+G_I(s))}.
    \label{eq:iq-sens}
\end{equation}
In \cite{qiu-imp}, they point out that the $(1+G_\phi(s))$ factor is cause for the larger unstable region under $A/\phi$ control. We now discuss how this factor makes a difference in terms of control.

\begin{figure}
    \centering
    \includegraphics[width=\linewidth]{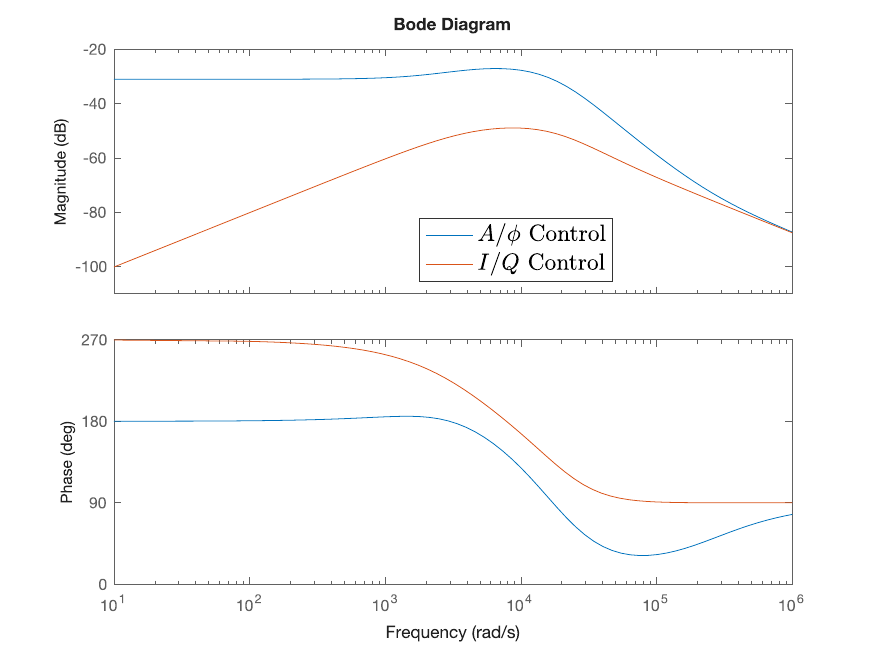}
    \caption{Bode plots for \cref{eq:sens-tf} and \cref{eq:iq-sens} when $\psi_0=22.5^\circ$.}
    \label{fig:ap-iq-bode}
\end{figure}

Comparing \cref{eq:sens-tf} and \cref{eq:iq-sens}, there are two striking differences. The first is that in $A/\phi$ control, the phase controller $(1+G_\phi(s))$ provides additional gain for translating $\dP\rightarrow\dV$. The second is that the amplitude controller $G_A(s)$ which would damp this effect is itself damped by $\cos{\psi_0}$. This effect is highlighted in when comparing the frequency response of the two transfer functions (\cref{fig:ap-iq-bode}). The ADRC controllers still provide a peak frequency in the response, but the magnitude is $\sim20$~dB less at this peak for $I/Q$ control. 

\begin{figure}
    \centering
    \includegraphics[width=\linewidth]{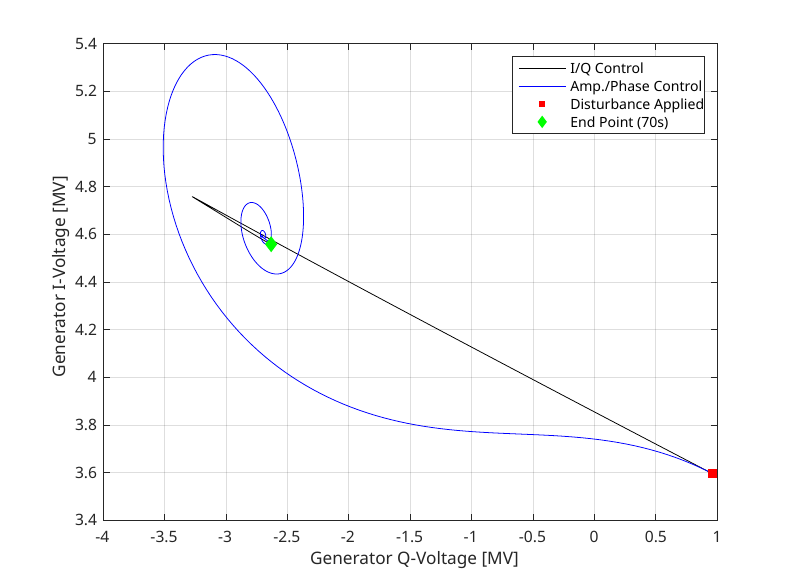}
    \caption{Generator Q-voltage v. generator I-voltage for $A/\phi$ and $I/Q$ controlled cavities when a disturbance is applied.}
    \label{fig:trajectory-comp}
\end{figure}

The result of these factors is that the path the generator output takes in correcting for disturbances is more sensitive to vibrations with $A/\phi$ control than with $I/Q$ control. We demonstrate this effect with two simulink models: one configured for $A/\phi$ control and the other for $I/Q$ control using ADRC controllers. We applied a $45^\circ$ ($\wh$) kick to the detuning and compare the generator outputs in \cref{fig:trajectory-comp}. The oscillations under $A/\phi$ control can be seen clearly in the looping path taken, whereas the $I/Q$ controlled system follows a more direct, linear path. 

\begin{figure}
    \centering
    \includegraphics[width=0.85\linewidth]{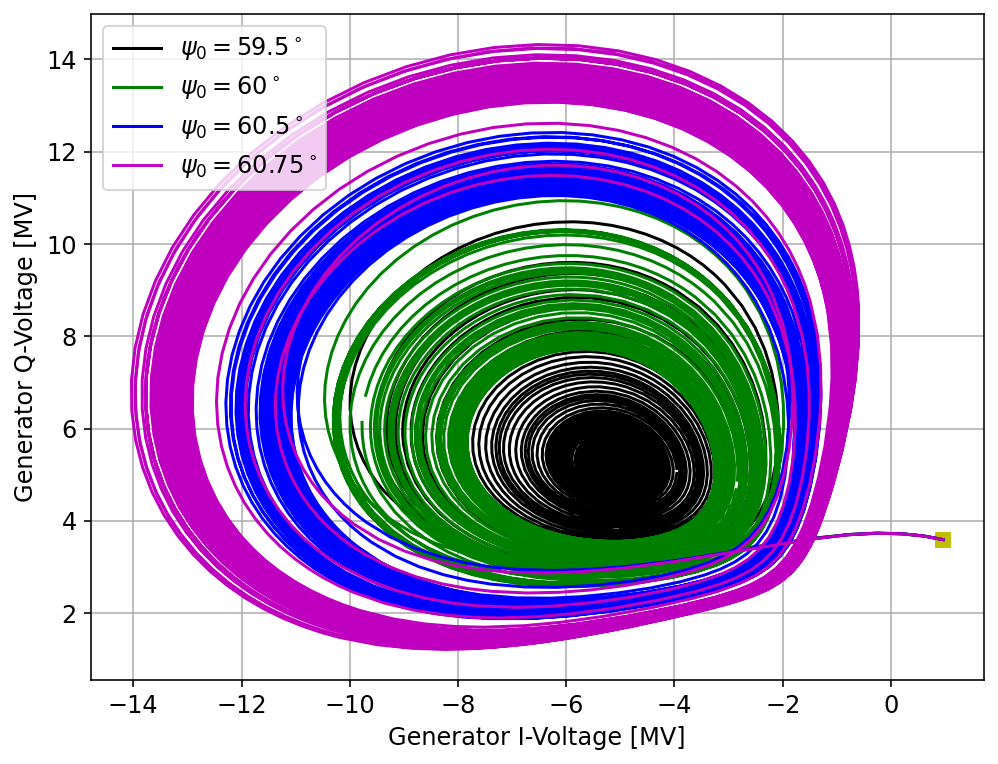}
    \caption{Generator Q-voltage v. generator I-voltage for a few $\psi_0$ near the positive side, ADRC instability threshold at $\psi_0=61^\circ$.}
    \label{fig:ivq-ap}
\end{figure}

This helps to explain how these crosstalk effects physically manifest in the controlled cavity. In the presence of ponderomotive oscillations, this effect can be exaggerated and lead to lower instability thresholds as the generator cannot provide the proper corrections to suppress them. Without strong crosstalk, as in $I/Q$ control, the generator is able to provide more direct corrections. This is also the mechanisms through which controller instability can occur in $A/\phi$ control. As the detuning increases and the amplitude controller is damped by $\cos{\psi_0}$, the crosstalk can cause the generator to never converge.
[floatfix]
This is the case in our Simulink model; as we detune the cavity up to just before the instability threshold, the generator can no longer converge to a steady working point. We present this effect in \cref{fig:ivq-ap}. Instead of converging to a stable working point, the generator output orbits in the complex plane with the controller response frequency. At the instability threshold ($\psi_0=61^\circ$), these orbits are no longer bounded and the generator output diverges to infinity. 

\subsubsection{Comparison with PI Control}
ADRC is a recent advent of the last 30 years, and has shown promise in implementation in industry and has been successful in the implementation for amplitude and phase control at FRIB. Historically however, proportional-integral (PI) control has shown success in the control of amplitude and phase of SRF cavities. In this section we seek to compare the two control schemes to each other in regards to preventing ponderomotive instabilities. For completeness, we note the relevant continuous transfer function for a standard PI controller:
\begin{equation}
    G_c(s) = k_P+\frac{k_I}{s},
    \label{eq:PI-controller}
\end{equation}
where $k_P$ is the proportional gain, and $k_I$ is the integral gain.

There is no easy way to make an equivalent PI controller given an ADRC control scheme; the control scheme in ADRC itself (ignoring the ESO) is PD \cite{1242516}, but this controller is not used in LLRF applications. In terms of comparison, we will start using gain sets reported in \cite{qiu-imp} and their interaction with our specific cavity. The work of \cite{qiu-imp} details an exhaustive creation of a reliable cavity simulator including the effects of solid state amplifiers, for which our simulation does not. To avoid having to compare two simulators (and for any others), we will keep this comparison to analytical means only. 

\begin{figure}
    \centering
    \includegraphics[width=\linewidth]{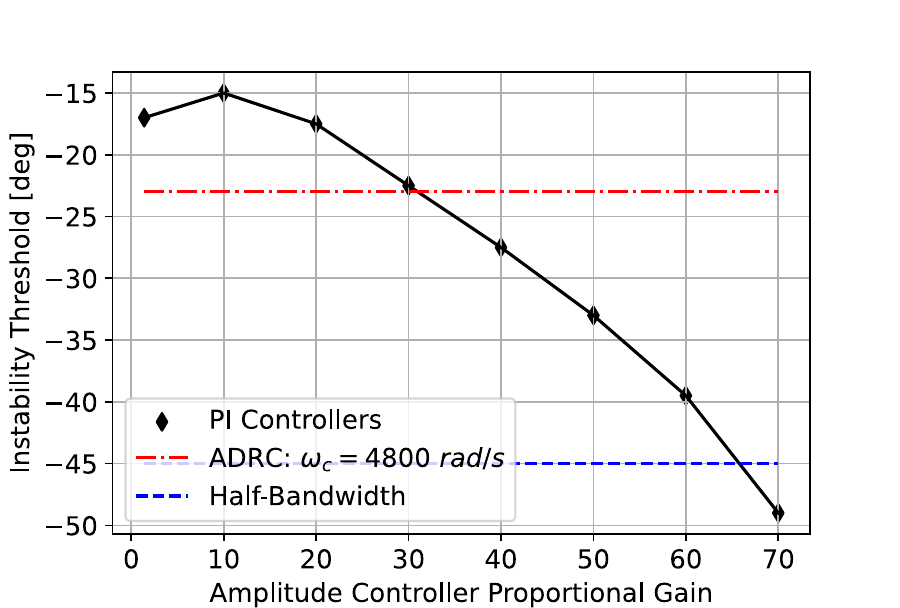}
    \caption{Instability threshold v. amplitude controller proportional gain $k_{AP}$.}
    \label{fig:pi-comparison}
\end{figure}

Other differences that are to be noted stem from an electrical and mechanical standpoint: FRIB HWRs are less loaded (higher $Q_L$, smaller bandwidths) than the CAFE2 HWR010 by $5-10$ times. Additionally, the specification voltages for both $\beta=0.29$ and $\beta=0.53$ HWRs are slightly higher for similar $k_\mu$. Using Table 4 in \cite{qiu-imp} and our analytical modeling, we find that $k_{AP}=1.4,~k_{AI}=200,~k_{\phi P}=10,~k_{\phi I}=1000$ produces the highest instability threshold of $\psi_0=-17^\circ$. We then varied the amplitude proportional gain until the threshold is pushed to $-45^\circ$, which we summarize in \cref{fig:pi-comparison}.

We see that for our $\beta=0.53$ HWRs, the amplitude proportional gain has to be increased about 30 times in order to match the operational ADRC LLRF parameters. The proportional gain needs to be increased almost 60 times in order to push the instability threshold to beyond the half-bandwidth.
[floatfix]
\section{Discussion}
In this work, we have presented an extensive analysis of ponderomotive effects in narrow-band, coaxial resonators. We have shown that even in self-excited loop mode, ponderomotive oscillations can occur for high-frequency mechanical modes. Self excited loop is generally considered better than generator driven mode when it comes to a reduction in ponderomotive effects \cite{delayen, DELAYEN20061}, so this phenomena is of great interest to us.

On the closed loop front, we have shown for non-linear and linearized ADRC that ponderomotive oscillations can couple to the ADRC frequency response for high amplitude controller bandwidths. This effectively lowers the oscillatory instability threshold by exciting higher harmonics of the mechanical resonance. We have shown that this effect requires higher frequency mechanical resonances; a 90 Hz mechanical resonance with the same $k_\mu$ and $Q_\mu$ does not experience this coupling. 

Both of studies suggest that there are more considerations that can be made regarding instabilities when designing SRF cavities. Conventional cavity design philosophy is to push the first mechanical resonance to above 100~Hz. This is good from a microphonics perspective; modes above 100~Hz are less likely to be excited by ambient noise. Our open and closed loop observations tell us that having a mechanical resonance above 100~Hz does not necessarily reduce the risk of the oscillatory instability however. 

In the future, we plan to examine the the oscillatory instability from a cavity design lens. This will give us a better insight to which mechanical and electrical parameters are useful in reducing the risk for the oscillatory instability. With analytical modeling, we can then validate that an optimized open loop cavity corresponds to an optimized closed loop cavity. 

We now return to the discussion about the positive side instability with ADRC. It is clear to us that the instability that we observe in the machine is linked to the ADRC controller, however there are still some unknowns. Our Simulink model is not a 1-to-1 recreation of the control algorithm deployed to the FRIB LLRF system. FRIB LLRF utilizes a phase-change limiter to avoid jumps in the drive phase from one quadrant to another as is common for $A/\phi$ control. This aspect is not included in the Simulink or analytical models, and could account for the difference we see when comparing the two to the accelerator measurements. Additionally, our Simulink model does not include safety interlock systems. At this time, it is hard to determine if the instability threshold is lower in the real machine due to the difference in deployed algorithms or a trip in the safety interlocks due to phenomena as seen in \cref{fig:ivq-ap}.

\section{Conclusions}
We have experimentally shown the presence of ponderomotive oscillations in self-excited, coaxial resonators at the STC at Fermilab. We have presented a preliminary analytical method to examine these instabilities. This lays the groundwork for a future work to consider these instabilities from a cavity design perspective.

Additionally, we have also examined the effect of non-linear and linearized ADRC on the oscillatory instability when used for $A/\phi$ control. We explain previous observations about the behavior of the instability threshold as a function of amplitude controller gain: high-frequency mechanical resonances can couple to the ADRC frequency response via higher harmonics. Our analytical and simulation model suggest that the instability we see on the positive side in accelerator cavities is a result of controller instability rather than ponderomotive effects.

Despite these complications, we show that ADRC out-preforms a PI controller in terms of instability threshold when using proportional and integral gains utilized for a similarly electromechanically coupled resonance. In this comparison, the proportional gain must be increased 30 times in order to provide the same instability threshold as our ADRC. To push the threshold beyond the half-bandwidth, we must use a proportional gain 60 times higher.

We reaffirm the findings of \cite{qiu-imp} with regards to $I/Q$ control: $I/Q$ control provides substantially higher instability thresholds compared to $A/\phi$ control. We expand on previous analysis, showing how increased crosstalk between amplitude and phase will lead to more convoluted corrections made by the generator. We also use our Simulink model to show how this effect can lead to controller instabilities not present in $I/Q$ control. 

\section{Acknowledgments}
In addition to the footnote source, this material is based upon work supported by the U.S. Department of Energy, Office of Science, Office of Nuclear Physics and used the resources of the Facility for Rare Isotope Beams (FRIB) operations, which is a DOE Office of Science User Facility under award number DE-SC0023633.

Additionally, this work was produced by Fermi Forward Discovery Group, LLC under Contract No. 89243024CSC000002 with the U.S. Department of Energy, Office of Science, Office of High Energy Physics. Publisher acknowledges the U.S. Government license to provide public access under the DOE Public Access Plan \url{https://www.energy.gov/downloads/doe-public-access-plan}.

\bibliography{main_bib}% Produces the bibliography via BibTeX.

%apsrev4-2.bst 2019-01-14 (MD) hand-edited version of apsrev4-1.bst
%Control: key (0)
%Control: author (8) initials jnrlst
%Control: editor formatted (1) identically to author
%Control: production of article title (0) allowed
%Control: page (0) single
%Control: year (1) truncated
%Control: production of eprint (0) enabled
\begin{thebibliography}{43}%
\makeatletter
\providecommand \@ifxundefined [1]{%
 \@ifx{#1\undefined}
}%
\providecommand \@ifnum [1]{%
 \ifnum #1\expandafter \@firstoftwo
 \else \expandafter \@secondoftwo
 \fi
}%
\providecommand \@ifx [1]{%
 \ifx #1\expandafter \@firstoftwo
 \else \expandafter \@secondoftwo
 \fi
}%
\providecommand \natexlab [1]{#1}%
\providecommand \enquote  [1]{``#1''}%
\providecommand \bibnamefont  [1]{#1}%
\providecommand \bibfnamefont [1]{#1}%
\providecommand \citenamefont [1]{#1}%
\providecommand \href@noop [0]{\@secondoftwo}%
\providecommand \href [0]{\begingroup \@sanitize@url \@href}%
\providecommand \@href[1]{\@@startlink{#1}\@@href}%
\providecommand \@@href[1]{\endgroup#1\@@endlink}%
\providecommand \@sanitize@url [0]{\catcode `\\12\catcode `\$12\catcode `\&12\catcode `\#12\catcode `\^12\catcode `\_12\catcode `\%12\relax}%
\providecommand \@@startlink[1]{}%
\providecommand \@@endlink[0]{}%
\providecommand \url  [0]{\begingroup\@sanitize@url \@url }%
\providecommand \@url [1]{\endgroup\@href {#1}{\urlprefix }}%
\providecommand \urlprefix  [0]{URL }%
\providecommand \Eprint [0]{\href }%
\providecommand \doibase [0]{https://doi.org/}%
\providecommand \selectlanguage [0]{\@gobble}%
\providecommand \bibinfo  [0]{\@secondoftwo}%
\providecommand \bibfield  [0]{\@secondoftwo}%
\providecommand \translation [1]{[#1]}%
\providecommand \BibitemOpen [0]{}%
\providecommand \bibitemStop [0]{}%
\providecommand \bibitemNoStop [0]{.\EOS\space}%
\providecommand \EOS [0]{\spacefactor3000\relax}%
\providecommand \BibitemShut  [1]{\csname bibitem#1\endcsname}%
\let\auto@bib@innerbib\@empty
%</preamble>
\bibitem [{\citenamefont {Fortuna}\ \emph {et~al.}(1990)\citenamefont {Fortuna} \emph {et~al.}}]{FORTUNA1990253}%
  \BibitemOpen
  \bibfield  {author} {\bibinfo {author} {\bibfnamefont {G.}~\bibnamefont {Fortuna}} \emph {et~al.},\ }\bibfield  {title} {\bibinfo {title} {The {ALPI} project at the {L}aboratori {N}azionali di {L}egnaro},\ }\href {https://doi.org/https://doi.org/10.1016/0168-9002(90)91803-J} {\bibfield  {journal} {\bibinfo  {journal} {Nuclear Instruments and Methods in Physics Research Section A: Accelerators, Spectrometers, Detectors and Associated Equipment}\ }\textbf {\bibinfo {volume} {287}},\ \bibinfo {pages} {253} (\bibinfo {year} {1990})}\BibitemShut {NoStop}%
\bibitem [{\citenamefont {Lab.}(1978)}]{osti_5991052}%
  \BibitemOpen
  \bibfield  {author} {\bibinfo {author} {\bibfnamefont {A.~N.}\ \bibnamefont {Lab.}},\ }\href {https://doi.org/10.2172/5991052} {\emph {\bibinfo {title} {ATLAS: a proposal for a precision heavy ion accelerator at Argonne National Laboratory}}},\ \bibinfo {type} {Tech. Rep.}\ (\bibinfo  {institution} {Argonne National Lab., IL (USA)},\ \bibinfo {year} {1978})\BibitemShut {NoStop}%
\bibitem [{\citenamefont {Kester}\ \emph {et~al.}(2009)\citenamefont {Kester} \emph {et~al.}}]{ReA}%
  \BibitemOpen
  \bibfield  {author} {\bibinfo {author} {\bibfnamefont {O.}~\bibnamefont {Kester}} \emph {et~al.},\ }\bibfield  {title} {\bibinfo {title} {The {MSU/NSCL} {R}e-{A}ccelerator {ReA3}},\ }in\ \href@noop {} {\emph {\bibinfo {booktitle} {Proc. of SRF2009: The 14th International Conference on RF Superconductivity}}}\ (\bibinfo {year} {2009})\BibitemShut {NoStop}%
\bibitem [{\citenamefont {Lewitowicz}(2008)}]{spiral2}%
  \BibitemOpen
  \bibfield  {author} {\bibinfo {author} {\bibfnamefont {M.}~\bibnamefont {Lewitowicz}},\ }\bibfield  {title} {\bibinfo {title} {The {SPIRAL 2} project},\ }\href {https://doi.org/https://doi.org/10.1016/j.nuclphysa.2008.02.290} {\bibfield  {journal} {\bibinfo  {journal} {Nuclear Physics A}\ }\textbf {\bibinfo {volume} {805}},\ \bibinfo {pages} {519c} (\bibinfo {year} {2008})},\ \bibinfo {note} {iNPC 2007}\BibitemShut {NoStop}%
\bibitem [{\citenamefont {Wei}\ \emph {et~al.}(2022)\citenamefont {Wei} \emph {et~al.}}]{mpla37:2230006}%
  \BibitemOpen
  \bibfield  {author} {\bibinfo {author} {\bibfnamefont {J.}~\bibnamefont {Wei}} \emph {et~al.},\ }\bibfield  {title} {\bibinfo {title} {Accelerator commissioning and rare isotope identification at the {Facility for Rare Isotope Beams}},\ }\href {https://doi.org/10.1142/S0217732322300063} {\bibfield  {journal} {\bibinfo  {journal} {Mod. Phys. Lett. A}\ }\textbf {\bibinfo {volume} {37}},\ \bibinfo {pages} {2230006} (\bibinfo {year} {2022})}\BibitemShut {NoStop}%
\bibitem [{\citenamefont {Wei}\ \emph {et~al.}(2023)\citenamefont {Wei} \emph {et~al.}}]{osti_og}%
  \BibitemOpen
  \bibfield  {author} {\bibinfo {author} {\bibfnamefont {J.}~\bibnamefont {Wei}} \emph {et~al.},\ }\bibfield  {title} {\bibinfo {title} {{FRIB} transition to user operations, power ramp up, and upgrade perspectives},\ }in\ \href {https://doi.org/10.18429/JACoW-SRF2023-MOIAA01} {\emph {\bibinfo {booktitle} {Proc. SRF 2023: 21st Int. Conf. RF Superconductivity, Grand Rapids, MI, USA}}}\ (\bibinfo  {publisher} {JACoW},\ \bibinfo {year} {2023})\ pp.\ \bibinfo {pages} {1--8}\BibitemShut {NoStop}%
\bibitem [{\citenamefont {Xu}\ \emph {et~al.}(2017)\citenamefont {Xu} \emph {et~al.}}]{Xu:SRF2017-TUXAA03}%
  \BibitemOpen
  \bibfield  {author} {\bibinfo {author} {\bibfnamefont {T.}~\bibnamefont {Xu}} \emph {et~al.},\ }\bibfield  {title} {{\selectlanguage {english}\bibinfo {title} {{P}rogress of {FRIB} {SRF} production}},\ }in\ \href {https://doi.org/https://doi.org/10.18429/JACoW-SRF2017-TUXAA03} {{\selectlanguage {english}\emph {\bibinfo {booktitle} {Proc. of 18'th International Conference on RF Superconductivity (SRF'17), Lanzhou, China, July 17-21}}}},\ \bibinfo {series and number} {\bibinfo {series} {International Conference on RF Superconductivity}\ No.~\bibinfo {number} {18}}\ (\bibinfo {year} {2017})\ pp.\ \bibinfo {pages} {345--352}\BibitemShut {NoStop}%
\bibitem [{\citenamefont {Adetunji}\ \emph {et~al.}(2025)\citenamefont {Adetunji} \emph {et~al.}}]{pip2}%
  \BibitemOpen
  \bibfield  {author} {\bibinfo {author} {\bibfnamefont {J.}~\bibnamefont {Adetunji}} \emph {et~al.} (\bibinfo {collaboration} {PIP-II}),\ }\href@noop {} {\emph {\bibinfo {title} {{Proton Improvement Plan-II Final Design Report}}}},\ \bibinfo {type} {Tech. Rep.}\ (\bibinfo  {institution} {Fermi National Accelerator Laboratory},\ \bibinfo {year} {2025})\BibitemShut {NoStop}%
\bibitem [{\citenamefont {R.~Mitchell}\ \emph {et~al.}(2001)\citenamefont {R.~Mitchell}, \citenamefont {Ciovati}, \citenamefont {Davis}, \citenamefont {Macha},\ and\ \citenamefont {Sundelin}}]{sns-lfd}%
  \BibitemOpen
  \bibfield  {author} {\bibinfo {author} {\bibfnamefont {K.~M.}\ \bibnamefont {R.~Mitchell}}, \bibinfo {author} {\bibfnamefont {G.}~\bibnamefont {Ciovati}}, \bibinfo {author} {\bibfnamefont {K.}~\bibnamefont {Davis}}, \bibinfo {author} {\bibfnamefont {K.}~\bibnamefont {Macha}},\ and\ \bibinfo {author} {\bibfnamefont {R.}~\bibnamefont {Sundelin}},\ }\bibfield  {title} {\bibinfo {title} {Lorentz force detuning analysis of the {S}pallation {N}eutron {S}ource ({SNS}) accelerating cavities},\ }in\ \href@noop {} {\emph {\bibinfo {booktitle} {Proc. of SRF2001: The 10th Workshop on RF Superconducivity}}}\ (\bibinfo {year} {2001})\BibitemShut {NoStop}%
\bibitem [{\citenamefont {Gonin}\ \emph {et~al.}(2016)\citenamefont {Gonin} \emph {et~al.}}]{stiffness-lfd}%
  \BibitemOpen
  \bibfield  {author} {\bibinfo {author} {\bibfnamefont {I.~V.}\ \bibnamefont {Gonin}} \emph {et~al.},\ }\bibfield  {title} {\bibinfo {title} {650 {MHz} elliptical superconducting {RF} cavities for {PIP-II} project},\ }in\ \href@noop {} {\emph {\bibinfo {booktitle} {Proc. of LINAC2016: the 28th Linear Accelerator Conference}}}\ (\bibinfo {year} {2016})\BibitemShut {NoStop}%
\bibitem [{\citenamefont {Schulze}(1971)}]{schulze}%
  \BibitemOpen
  \bibfield  {author} {\bibinfo {author} {\bibfnamefont {D.}~\bibnamefont {Schulze}},\ }\emph {\bibinfo {title} {Ponderomotive stability of RF resonators and resonator control systems}},\ \href@noop {} {Ph.D. thesis},\ \bibinfo  {school} {Inst. fuer Experimentelle Kernphysik, Kernforschungszentrum Karlsruhe (F.R. Germany).} (\bibinfo {year} {1971})\BibitemShut {NoStop}%
\bibitem [{\citenamefont {Delayen}(1978)}]{delayen}%
  \BibitemOpen
  \bibfield  {author} {\bibinfo {author} {\bibfnamefont {J.~R.}\ \bibnamefont {Delayen}},\ }\emph {\bibinfo {title} {Phase and amplitude stabilization of superconducting resonators}},\ \href@noop {} {Ph.D. thesis},\ \bibinfo  {school} {California Institute of Technology} (\bibinfo {year} {1978})\BibitemShut {NoStop}%
\bibitem [{\citenamefont {Sukhanov}\ \emph {et~al.}(2019)\citenamefont {Sukhanov} \emph {et~al.}}]{stc}%
  \BibitemOpen
  \bibfield  {author} {\bibinfo {author} {\bibfnamefont {A.}~\bibnamefont {Sukhanov}} \emph {et~al.},\ }\bibfield  {title} {\bibinfo {title} {Upgrade of the {F}ermilab {S}poke {T}est {C}ryostat for testing of {PIP-II} 650 {MHz} 5-cell ellipticals},\ }in\ \href@noop {} {\emph {\bibinfo {booktitle} {Proc. of 19th International Conference on RF Superconductivity}}}\ (\bibinfo {year} {2019})\BibitemShut {NoStop}%
\bibitem [{\citenamefont {Sukhanov}\ \emph {et~al.}(2024)\citenamefont {Sukhanov}, \citenamefont {Contreras-Martinez}, \citenamefont {Grimm}, \citenamefont {Hanna}, \citenamefont {Hansen}, \citenamefont {Kazakov}, \citenamefont {Khabiboulline}, \citenamefont {Parise}, \citenamefont {Passarelli}, \citenamefont {Pischalnikov} \emph {et~al.}}]{stc2}%
  \BibitemOpen
  \bibfield  {author} {\bibinfo {author} {\bibfnamefont {A.}~\bibnamefont {Sukhanov}}, \bibinfo {author} {\bibfnamefont {C.}~\bibnamefont {Contreras-Martinez}}, \bibinfo {author} {\bibfnamefont {C.}~\bibnamefont {Grimm}}, \bibinfo {author} {\bibfnamefont {B.}~\bibnamefont {Hanna}}, \bibinfo {author} {\bibfnamefont {B.}~\bibnamefont {Hansen}}, \bibinfo {author} {\bibfnamefont {S.}~\bibnamefont {Kazakov}}, \bibinfo {author} {\bibfnamefont {T.}~\bibnamefont {Khabiboulline}}, \bibinfo {author} {\bibfnamefont {M.}~\bibnamefont {Parise}}, \bibinfo {author} {\bibfnamefont {D.}~\bibnamefont {Passarelli}}, \bibinfo {author} {\bibfnamefont {Y.}~\bibnamefont {Pischalnikov}}, \emph {et~al.},\ }\bibfield  {title} {\bibinfo {title} {Cold test results of pre-production {PIP-II} {SSR2} cavities with high-power coupler in the {F}ermilab {S}poke {T}est {C}ryostat}\ }(\bibinfo {organization} {Fermi National Accelerator Laboratory (FNAL), Batavia, IL (United States)},\ \bibinfo {year} {2024})\BibitemShut {NoStop}%
\bibitem [{\citenamefont {Pischalnikov}\ \emph {et~al.}(2010)\citenamefont {Pischalnikov}, \citenamefont {Carcagno}, \citenamefont {Makulski}, \citenamefont {Orris},\ and\ \citenamefont {Schappert}}]{osti_1969329}%
  \BibitemOpen
  \bibfield  {author} {\bibinfo {author} {\bibfnamefont {Y.}~\bibnamefont {Pischalnikov}}, \bibinfo {author} {\bibfnamefont {R.}~\bibnamefont {Carcagno}}, \bibinfo {author} {\bibfnamefont {A.}~\bibnamefont {Makulski}}, \bibinfo {author} {\bibfnamefont {D.}~\bibnamefont {Orris}},\ and\ \bibinfo {author} {\bibfnamefont {W.}~\bibnamefont {Schappert}},\ }\bibfield  {title} {\bibinfo {title} {Development of {SCRF} cavity resonance control algorithms at {Fermilab}},\ }\href {https://www.osti.gov/biblio/1969329} {\bibfield  {journal} {\bibinfo  {journal} {TBD}\ } (\bibinfo {year} {2010})}\BibitemShut {NoStop}%
\bibitem [{\citenamefont {Schappert}\ \emph {et~al.}(2011)\citenamefont {Schappert}, \citenamefont {Pischalnikov},\ and\ \citenamefont {Scorrano}}]{Schappert:2011zz}%
  \BibitemOpen
  \bibfield  {author} {\bibinfo {author} {\bibfnamefont {W.}~\bibnamefont {Schappert}}, \bibinfo {author} {\bibfnamefont {Y.}~\bibnamefont {Pischalnikov}},\ and\ \bibinfo {author} {\bibfnamefont {M.}~\bibnamefont {Scorrano}} (\bibinfo {collaboration} {Project X}),\ }\bibfield  {title} {\bibinfo {title} {{Resonance Control in {SRF} Cavities at {FNAL}}},\ }\href@noop {} {\bibfield  {journal} {\bibinfo  {journal} {Conf. Proc. C}\ }\textbf {\bibinfo {volume} {110328}},\ \bibinfo {pages} {2130} (\bibinfo {year} {2011})}\BibitemShut {NoStop}%
\bibitem [{\citenamefont {Pischalnikov}\ \emph {et~al.}(2012)\citenamefont {Pischalnikov}, \citenamefont {Cancelo}, \citenamefont {Chase}, \citenamefont {Crawford}, \citenamefont {Edstrom}, \citenamefont {Harms}, \citenamefont {Costin}, \citenamefont {Solyak},\ and\ \citenamefont {Schappert}}]{pischalnikov-THPPR012}%
  \BibitemOpen
  \bibfield  {author} {\bibinfo {author} {\bibfnamefont {Y.}~\bibnamefont {Pischalnikov}}, \bibinfo {author} {\bibfnamefont {G.}~\bibnamefont {Cancelo}}, \bibinfo {author} {\bibfnamefont {B.}~\bibnamefont {Chase}}, \bibinfo {author} {\bibfnamefont {D.}~\bibnamefont {Crawford}}, \bibinfo {author} {\bibfnamefont {D.}~\bibnamefont {Edstrom}}, \bibinfo {author} {\bibfnamefont {E.}~\bibnamefont {Harms}}, \bibinfo {author} {\bibfnamefont {R.}~\bibnamefont {Costin}}, \bibinfo {author} {\bibfnamefont {N.}~\bibnamefont {Solyak}},\ and\ \bibinfo {author} {\bibfnamefont {W.}~\bibnamefont {Schappert}},\ }\bibfield  {title} {\bibinfo {title} {Lorentz force compensation for long pulses in {SRF} cavities},\ }in\ \href@noop {} {\emph {\bibinfo {booktitle} {Proc. of 3rd International Particle Accelerator Conference}}}\ (\bibinfo  {publisher} {JACoW},\ \bibinfo {year} {2012})\ pp.\ \bibinfo {pages} {3990--3992}\BibitemShut {NoStop}%
\bibitem [{\citenamefont {Schappert}\ \emph {et~al.}(2015)\citenamefont {Schappert}, \citenamefont {Holzbauer},\ and\ \citenamefont {Pischalnikov}}]{Schappert:IPAC2015-WEPTY036}%
  \BibitemOpen
  \bibfield  {author} {\bibinfo {author} {\bibfnamefont {W.}~\bibnamefont {Schappert}}, \bibinfo {author} {\bibfnamefont {J.}~\bibnamefont {Holzbauer}},\ and\ \bibinfo {author} {\bibfnamefont {Y.}~\bibnamefont {Pischalnikov}},\ }\bibfield  {title} {{\selectlanguage {english}\bibinfo {title} {{P}rogress at {FNAL} in the field of the active resonance control for narrow bandwidth {SRF} cavities.}},\ }in\ \href {https://doi.org/https://doi.org/10.18429/JACoW-IPAC2015-WEPTY036} {{\selectlanguage {english}\emph {\bibinfo {booktitle} {Proc. 6th International Particle Accelerator Conference (IPAC'15), Richmond, VA, USA, May 3-8, 2015}}}},\ \bibinfo {series and number} {\bibinfo {series} {International Particle Accelerator Conference}\ No.~\bibinfo {number} {6}}\ (\bibinfo  {publisher} {JACoW},\ \bibinfo {address} {Geneva, Switzerland},\ \bibinfo {year} {2015})\ pp.\ \bibinfo {pages} {3355--3357},\ \bibinfo {note} {https://doi.org/10.18429/JACoW-IPAC2015-WEPTY036}\BibitemShut {NoStop}%
\bibitem [{\citenamefont {Pfeiffer}\ \emph {et~al.}(2020)\citenamefont {Pfeiffer}, \citenamefont {Eichler},\ and\ \citenamefont {Schlarb}}]{PFEIFFER20201331}%
  \BibitemOpen
  \bibfield  {author} {\bibinfo {author} {\bibfnamefont {S.}~\bibnamefont {Pfeiffer}}, \bibinfo {author} {\bibfnamefont {A.}~\bibnamefont {Eichler}},\ and\ \bibinfo {author} {\bibfnamefont {H.}~\bibnamefont {Schlarb}},\ }\bibfield  {title} {\bibinfo {title} {Model-based feed-forward control for time-varying systems with an example for {SRF} cavities},\ }\href {https://doi.org/https://doi.org/10.1016/j.ifacol.2020.12.1868} {\bibfield  {journal} {\bibinfo  {journal} {IFAC-PapersOnLine}\ }\textbf {\bibinfo {volume} {53}},\ \bibinfo {pages} {1331} (\bibinfo {year} {2020})},\ \bibinfo {note} {21st IFAC World Congress}\BibitemShut {NoStop}%
\bibitem [{\citenamefont {Brown}\ \emph {et~al.}(2025)\citenamefont {Brown}, \citenamefont {Kim}, \citenamefont {Zhao}, \citenamefont {Xu},\ and\ \citenamefont {Chang}}]{brown:hiat2025-wep19}%
  \BibitemOpen
  \bibfield  {author} {\bibinfo {author} {\bibfnamefont {J.}~\bibnamefont {Brown}}, \bibinfo {author} {\bibfnamefont {S.}~\bibnamefont {Kim}}, \bibinfo {author} {\bibfnamefont {S.}~\bibnamefont {Zhao}}, \bibinfo {author} {\bibfnamefont {T.}~\bibnamefont {Xu}},\ and\ \bibinfo {author} {\bibfnamefont {W.}~\bibnamefont {Chang}},\ }\bibfield  {title} {{\selectlanguage {English}\bibinfo {title} {Study of ponderomotive instability in the {FRIB} beta=0.53 half-wave resonator}},\ }in\ \href {https://doi.org/10.18429/JACoW-HIAT2025-WEP19} {{\selectlanguage {English}\emph {\bibinfo {booktitle} {Proc. 16th International Conference on Heavy Ion Accelerator Technology}}}},\ \bibinfo {series and number} {\bibinfo {series} {HIAT}\ No.~\bibinfo {number} {16}}\ (\bibinfo  {publisher} {JACoW Publishing, Geneva, Switzerland},\ \bibinfo {year} {2025})\ pp.\ \bibinfo {pages} {258--261}\BibitemShut {NoStop}%
\bibitem [{\citenamefont {Karliner}\ \emph {et~al.}(1967)\citenamefont {Karliner}, \citenamefont {Shapiro},\ and\ \citenamefont {Shekhtman}}]{karliner_instability_1967}%
  \BibitemOpen
  \bibfield  {author} {\bibinfo {author} {\bibfnamefont {M.}~\bibnamefont {Karliner}}, \bibinfo {author} {\bibfnamefont {V.}~\bibnamefont {Shapiro}},\ and\ \bibinfo {author} {\bibfnamefont {I.}~\bibnamefont {Shekhtman}},\ }\bibfield  {title} {\bibinfo {title} {Instability in the walls of a cavity due to ponderomotive forces of the electromagnetic field},\ }\href@noop {} {\bibfield  {journal} {\bibinfo  {journal} {Soviet Physics-Technical Physics}\ }\textbf {\bibinfo {volume} {11}},\ \bibinfo {pages} {1501} (\bibinfo {year} {1967})}\BibitemShut {NoStop}%
\bibitem [{\citenamefont {Karliner}\ \emph {et~al.}(1970)\citenamefont {Karliner}, \citenamefont {Petrov},\ and\ \citenamefont {Shekhtman}}]{karliner_vibration_1970}%
  \BibitemOpen
  \bibfield  {author} {\bibinfo {author} {\bibfnamefont {M.}~\bibnamefont {Karliner}}, \bibinfo {author} {\bibfnamefont {V.}~\bibnamefont {Petrov}},\ and\ \bibinfo {author} {\bibfnamefont {I.}~\bibnamefont {Shekhtman}},\ }\bibfield  {title} {\bibinfo {title} {Vibration of the walls of a cavity resonator under ponderomotive forces in the presence of feedback},\ }\href@noop {} {\bibfield  {journal} {\bibinfo  {journal} {Soviet Physics-Technical Physics}\ }\textbf {\bibinfo {volume} {14}},\ \bibinfo {pages} {1041} (\bibinfo {year} {1970})}\BibitemShut {NoStop}%
\bibitem [{\citenamefont {Delayen}(2006)}]{DELAYEN20061}%
  \BibitemOpen
  \bibfield  {author} {\bibinfo {author} {\bibfnamefont {J.}~\bibnamefont {Delayen}},\ }\bibfield  {title} {\bibinfo {title} {Ponderomotive instabilities and microphonics—a tutorial},\ }\href {https://doi.org/https://doi.org/10.1016/j.physc.2006.03.050} {\bibfield  {journal} {\bibinfo  {journal} {Physica C: Superconductivity}\ }\textbf {\bibinfo {volume} {441}},\ \bibinfo {pages} {1} (\bibinfo {year} {2006})},\ \bibinfo {note} {proceedings of the 12th International Workshop on RF Superconductivity}\BibitemShut {NoStop}%
\bibitem [{\citenamefont {Koscielniak}(2019{\natexlab{a}})}]{Koscielniak:IPAC2019-THPRB010}%
  \BibitemOpen
  \bibfield  {author} {\bibinfo {author} {\bibfnamefont {S.}~\bibnamefont {Koscielniak}},\ }\bibfield  {title} {{\selectlanguage {english}\bibinfo {title} {{P}onderomotive instability of generator driven cavity}},\ }in\ \href {https://doi.org/doi:10.18429/JACoW-IPAC2019-THPRB010} {{\selectlanguage {english}\emph {\bibinfo {booktitle} {Proc. 10th International Particle Accelerator Conference (IPAC'19), Melbourne, Australia, 19-24 May 2019}}}},\ \bibinfo {series and number} {\bibinfo {series} {International Particle Accelerator Conference}\ No.~\bibinfo {number} {10}}\ (\bibinfo  {publisher} {JACoW Publishing},\ \bibinfo {address} {Geneva, Switzerland},\ \bibinfo {year} {2019})\ pp.\ \bibinfo {pages} {3820--3823}\BibitemShut {NoStop}%
\bibitem [{\citenamefont {Koscielniak}(2019{\natexlab{b}})}]{Koscielniak:IPAC2019-THPRB009}%
  \BibitemOpen
  \bibfield  {author} {\bibinfo {author} {\bibfnamefont {S.}~\bibnamefont {Koscielniak}},\ }\bibfield  {title} {{\selectlanguage {english}\bibinfo {title} {{V}ector sum \& diffference control of {SRF} cavities}},\ }in\ \href {https://doi.org/doi:10.18429/JACoW-IPAC2019-THPRB009} {{\selectlanguage {english}\emph {\bibinfo {booktitle} {Proc. 10th International Particle Accelerator Conference (IPAC'19), Melbourne, Australia, 19-24 May 2019}}}},\ \bibinfo {series and number} {\bibinfo {series} {International Particle Accelerator Conference}\ No.~\bibinfo {number} {10}}\ (\bibinfo  {publisher} {JACoW Publishing},\ \bibinfo {address} {Geneva, Switzerland},\ \bibinfo {year} {2019})\ pp.\ \bibinfo {pages} {3816--3819}\BibitemShut {NoStop}%
\bibitem [{\citenamefont {Fong}\ and\ \citenamefont {Leewe}(2019)}]{Fong:IPAC2019-WEPRB003}%
  \BibitemOpen
  \bibfield  {author} {\bibinfo {author} {\bibfnamefont {K.}~\bibnamefont {Fong}}\ and\ \bibinfo {author} {\bibfnamefont {R.}~\bibnamefont {Leewe}},\ }\bibfield  {title} {{\selectlanguage {english}\bibinfo {title} {{P}arametric pumped oscillation by {L}orentz force in superconducting {RF} cavity}},\ }in\ \href {https://doi.org/doi:10.18429/JACoW-IPAC2019-WEPRB003} {{\selectlanguage {english}\emph {\bibinfo {booktitle} {Proc. 10th International Particle Accelerator Conference (IPAC'19), Melbourne, Australia, 19-24 May 2019}}}},\ \bibinfo {series and number} {\bibinfo {series} {International Particle Accelerator Conference}\ No.~\bibinfo {number} {10}}\ (\bibinfo  {publisher} {JACoW Publishing},\ \bibinfo {address} {Geneva, Switzerland},\ \bibinfo {year} {2019})\ pp.\ \bibinfo {pages} {2798--2800}\BibitemShut {NoStop}%
\bibitem [{\citenamefont {Bellandi}\ \emph {et~al.}(2020)\citenamefont {Bellandi}, \citenamefont {Branlard}, \citenamefont {Eichler},\ and\ \citenamefont {Pfeiffer}}]{BELLANDI2020361}%
  \BibitemOpen
  \bibfield  {author} {\bibinfo {author} {\bibfnamefont {A.}~\bibnamefont {Bellandi}}, \bibinfo {author} {\bibfnamefont {J.}~\bibnamefont {Branlard}}, \bibinfo {author} {\bibfnamefont {A.}~\bibnamefont {Eichler}},\ and\ \bibinfo {author} {\bibfnamefont {S.}~\bibnamefont {Pfeiffer}},\ }\bibfield  {title} {\bibinfo {title} {Integral resonance control in continuous wave superconducting particle accelerators},\ }\href {https://doi.org/https://doi.org/10.1016/j.ifacol.2020.12.186} {\bibfield  {journal} {\bibinfo  {journal} {IFAC-PapersOnLine}\ }\textbf {\bibinfo {volume} {53}},\ \bibinfo {pages} {361} (\bibinfo {year} {2020})},\ \bibinfo {note} {21st IFAC World Congress}\BibitemShut {NoStop}%
\bibitem [{\citenamefont {Bellandi}(2021)}]{bellandi-thesis}%
  \BibitemOpen
  \bibfield  {author} {\bibinfo {author} {\bibfnamefont {A.}~\bibnamefont {Bellandi}},\ }\emph {\bibinfo {title} {{LLRF} Contorl Techniques for the {E}uropean {XFEL} Continuous Wave Update}},\ \href@noop {} {Ph.D. thesis},\ \bibinfo  {school} {Universität Hamburg} (\bibinfo {year} {2021})\BibitemShut {NoStop}%
\bibitem [{\citenamefont {Qiu}\ \emph {et~al.}(2025)\citenamefont {Qiu}, \citenamefont {Peng}, \citenamefont {Wei}, \citenamefont {Miao}, \citenamefont {Xue}, \citenamefont {Zeng}, \citenamefont {Zhu}, \citenamefont {Jiang}, \citenamefont {Huang}, \citenamefont {Gao}, \citenamefont {Ma}, \citenamefont {Xu}, \citenamefont {Yang}, \citenamefont {Chen}, \citenamefont {Hou}, \citenamefont {Yang}, \citenamefont {Sun}, \citenamefont {Wang},\ and\ \citenamefont {He}}]{qiu-imp}%
  \BibitemOpen
  \bibfield  {author} {\bibinfo {author} {\bibfnamefont {F.}~\bibnamefont {Qiu}}, \bibinfo {author} {\bibfnamefont {J.}~\bibnamefont {Peng}}, \bibinfo {author} {\bibfnamefont {S.}~\bibnamefont {Wei}}, \bibinfo {author} {\bibfnamefont {Y.}~\bibnamefont {Miao}}, \bibinfo {author} {\bibfnamefont {Z.}~\bibnamefont {Xue}}, \bibinfo {author} {\bibfnamefont {R.}~\bibnamefont {Zeng}}, \bibinfo {author} {\bibfnamefont {Z.}~\bibnamefont {Zhu}}, \bibinfo {author} {\bibfnamefont {T.}~\bibnamefont {Jiang}}, \bibinfo {author} {\bibfnamefont {G.}~\bibnamefont {Huang}}, \bibinfo {author} {\bibfnamefont {Z.}~\bibnamefont {Gao}}, \bibinfo {author} {\bibfnamefont {J.}~\bibnamefont {Ma}}, \bibinfo {author} {\bibfnamefont {C.}~\bibnamefont {Xu}}, \bibinfo {author} {\bibfnamefont {L.}~\bibnamefont {Yang}}, \bibinfo {author} {\bibfnamefont {Z.}~\bibnamefont {Chen}}, \bibinfo {author} {\bibfnamefont {Q.}~\bibnamefont {Hou}}, \bibinfo {author} {\bibfnamefont {Z.}~\bibnamefont {Yang}}, \bibinfo {author} {\bibfnamefont
  {L.}~\bibnamefont {Sun}}, \bibinfo {author} {\bibfnamefont {Z.}~\bibnamefont {Wang}},\ and\ \bibinfo {author} {\bibfnamefont {Y.}~\bibnamefont {He}},\ }\bibfield  {title} {\bibinfo {title} {Analytical modeling and experimental verification of oscillatory ponderomotive instabilities in closed-loop superconducting {RF} cavities},\ }\href {https://doi.org/10.1103/hs5q-q5xc} {\bibfield  {journal} {\bibinfo  {journal} {Phys. Rev. Accel. Beams}\ }\textbf {\bibinfo {volume} {28}},\ \bibinfo {pages} {112001} (\bibinfo {year} {2025})}\BibitemShut {NoStop}%
\bibitem [{\citenamefont {Varghese}\ \emph {et~al.}(2023)\citenamefont {Varghese}, \citenamefont {Chase}, \citenamefont {Cullerton}, \citenamefont {Raman}, \citenamefont {Ahmed}, \citenamefont {Hanlet},\ and\ \citenamefont {Klepec}}]{varghese_llrf_2023}%
  \BibitemOpen
  \bibfield  {author} {\bibinfo {author} {\bibfnamefont {P.}~\bibnamefont {Varghese}}, \bibinfo {author} {\bibfnamefont {B.}~\bibnamefont {Chase}}, \bibinfo {author} {\bibfnamefont {E.}~\bibnamefont {Cullerton}}, \bibinfo {author} {\bibfnamefont {S.}~\bibnamefont {Raman}}, \bibinfo {author} {\bibfnamefont {S.}~\bibnamefont {Ahmed}}, \bibinfo {author} {\bibfnamefont {P.}~\bibnamefont {Hanlet}},\ and\ \bibinfo {author} {\bibfnamefont {D.}~\bibnamefont {Klepec}},\ }\href {https://doi.org/10.48550/arXiv.2311.00900} {\bibinfo {title} {{LLRF} system for the {Fermilab} {PIP}-{II} superconducting {LINAC}}} (\bibinfo {year} {2023}),\ \bibinfo {note} {arXiv:2311.00900 [physics]}\BibitemShut {NoStop}%
\bibitem [{\citenamefont {Doolittle}\ \emph {et~al.}(2017)\citenamefont {Doolittle} \emph {et~al.}}]{doolittle-lcls2}%
  \BibitemOpen
  \bibfield  {author} {\bibinfo {author} {\bibfnamefont {L.}~\bibnamefont {Doolittle}} \emph {et~al.},\ }\bibfield  {title} {\bibinfo {title} {High precision {RF} control for {SRF} cavities in lcls-ii},\ }in\ \href@noop {} {\emph {\bibinfo {booktitle} {Proc. of SRF2017: the 18th International Conference on RF Superconductivity}}}\ (\bibinfo {year} {2017})\BibitemShut {NoStop}%
\bibitem [{\citenamefont {Ma}\ \emph {et~al.}(2006)\citenamefont {Ma}, \citenamefont {Champion}, \citenamefont {Crofford}, \citenamefont {Kasemir}, \citenamefont {Piller}, \citenamefont {Doolittle},\ and\ \citenamefont {Ratti}}]{sns-llrf}%
  \BibitemOpen
  \bibfield  {author} {\bibinfo {author} {\bibfnamefont {H.}~\bibnamefont {Ma}}, \bibinfo {author} {\bibfnamefont {M.}~\bibnamefont {Champion}}, \bibinfo {author} {\bibfnamefont {M.}~\bibnamefont {Crofford}}, \bibinfo {author} {\bibfnamefont {K.-U.}\ \bibnamefont {Kasemir}}, \bibinfo {author} {\bibfnamefont {M.}~\bibnamefont {Piller}}, \bibinfo {author} {\bibfnamefont {L.}~\bibnamefont {Doolittle}},\ and\ \bibinfo {author} {\bibfnamefont {A.}~\bibnamefont {Ratti}},\ }\bibfield  {title} {\bibinfo {title} {Low-level {RF} control of {S}pallation {N}eutron {S}ource: System and characterization},\ }\href {https://doi.org/10.1103/PhysRevSTAB.9.032001} {\bibfield  {journal} {\bibinfo  {journal} {Phys. Rev. ST Accel. Beams}\ }\textbf {\bibinfo {volume} {9}},\ \bibinfo {pages} {032001} (\bibinfo {year} {2006})}\BibitemShut {NoStop}%
\bibitem [{\citenamefont {Ma}\ \emph {et~al.}(2024)\citenamefont {Ma} \emph {et~al.}}]{ma_virtual_2024}%
  \BibitemOpen
  \bibfield  {author} {\bibinfo {author} {\bibfnamefont {J.}~\bibnamefont {Ma}} \emph {et~al.},\ }\bibfield  {title} {\bibinfo {title} {Virtual cavity probe for the real-time identification of cavity burst-noise type in superconducting radio-frequency systems},\ }\href {https://doi.org/10.1016/j.nima.2024.169786} {\bibfield  {journal} {\bibinfo  {journal} {Nuclear Instruments and Methods in Physics Research Section A: Accelerators, Spectrometers, Detectors and Associated Equipment}\ }\textbf {\bibinfo {volume} {1068}},\ \bibinfo {pages} {169786} (\bibinfo {year} {2024})}\BibitemShut {NoStop}%
\bibitem [{\citenamefont {Bouly}\ \emph {et~al.}(2022)\citenamefont {Bouly}, \citenamefont {Giacomo}, \citenamefont {Leyge},\ and\ \citenamefont {Tontayeva}}]{bouly_superconducting_2022}%
  \BibitemOpen
  \bibfield  {author} {\bibinfo {author} {\bibfnamefont {F.}~\bibnamefont {Bouly}}, \bibinfo {author} {\bibfnamefont {M.~D.}\ \bibnamefont {Giacomo}}, \bibinfo {author} {\bibfnamefont {J.~F.}\ \bibnamefont {Leyge}},\ and\ \bibinfo {author} {\bibfnamefont {M.}~\bibnamefont {Tontayeva}},\ }\bibfield  {title} {\bibinfo {title} {Superconducting cavity and {RF} control loop model for the {SPIRAL2} linac},\ }\href {https://doi.org/10.18429/JACOW-LINAC2022-THPOPA13} {\bibfield  {journal} {\bibinfo  {journal} {Proc. of the 31st International Linear Accelerator Conference}\ }\textbf {\bibinfo {volume} {LINAC2022}},\ \bibinfo {pages} {pp. 767} (\bibinfo {year} {2022})}\BibitemShut {NoStop}%
\bibitem [{\citenamefont {Zhao}\ \emph {et~al.}(2013)\citenamefont {Zhao}, \citenamefont {Usher}, \citenamefont {Morris},\ and\ \citenamefont {Vincent}}]{zhao-fp}%
  \BibitemOpen
  \bibfield  {author} {\bibinfo {author} {\bibfnamefont {S.}~\bibnamefont {Zhao}}, \bibinfo {author} {\bibfnamefont {N.}~\bibnamefont {Usher}}, \bibinfo {author} {\bibfnamefont {D.}~\bibnamefont {Morris}},\ and\ \bibinfo {author} {\bibfnamefont {J.}~\bibnamefont {Vincent}},\ }\bibfield  {title} {\bibinfo {title} {Fixed-point implementation of active disturbance rejection control for superconducting radio frequency cavities},\ }in\ \href {https://doi.org/10.1109/ACC.2013.6580241} {\emph {\bibinfo {booktitle} {2013 American Control Conference}}}\ (\bibinfo {year} {2013})\ pp.\ \bibinfo {pages} {2693--2698}\BibitemShut {NoStop}%
\bibitem [{\citenamefont {Han}(2009)}]{han-trans}%
  \BibitemOpen
  \bibfield  {author} {\bibinfo {author} {\bibfnamefont {J.}~\bibnamefont {Han}},\ }\bibfield  {title} {\bibinfo {title} {From {PID} to active disturbance rejection control},\ }\href {https://doi.org/10.1109/TIE.2008.2011621} {\bibfield  {journal} {\bibinfo  {journal} {IEEE Transactions on Industrial Electronics}\ }\textbf {\bibinfo {volume} {56}},\ \bibinfo {pages} {900} (\bibinfo {year} {2009})}\BibitemShut {NoStop}%
\bibitem [{\citenamefont {Miklosovic}\ \emph {et~al.}(2006)\citenamefont {Miklosovic}, \citenamefont {Radke},\ and\ \citenamefont {Gao}}]{disc-ESO}%
  \BibitemOpen
  \bibfield  {author} {\bibinfo {author} {\bibfnamefont {R.}~\bibnamefont {Miklosovic}}, \bibinfo {author} {\bibfnamefont {A.}~\bibnamefont {Radke}},\ and\ \bibinfo {author} {\bibfnamefont {Z.}~\bibnamefont {Gao}},\ }\bibfield  {title} {\bibinfo {title} {Discrete implementation and generalization of the extended state observer},\ }in\ \href {https://doi.org/10.1109/ACC.2006.1656547} {\emph {\bibinfo {booktitle} {2006 American Control Conference}}}\ (\bibinfo {year} {2006})\ pp.\ \bibinfo {pages} {6--}\BibitemShut {NoStop}%
\bibitem [{\citenamefont {Zhao}\ and\ \citenamefont {Gao}(2010)}]{zhao-nmp}%
  \BibitemOpen
  \bibfield  {author} {\bibinfo {author} {\bibfnamefont {S.}~\bibnamefont {Zhao}}\ and\ \bibinfo {author} {\bibfnamefont {Z.}~\bibnamefont {Gao}},\ }\bibfield  {title} {\bibinfo {title} {Active disturbance rejection control for non-minimum phase systems},\ }in\ \href@noop {} {\emph {\bibinfo {booktitle} {Proceedings of the 29th Chinese Control Conference}}}\ (\bibinfo {year} {2010})\ pp.\ \bibinfo {pages} {6066--6070}\BibitemShut {NoStop}%
\bibitem [{\citenamefont {Tian}\ and\ \citenamefont {Gao}(2007)}]{tian-gao-tf}%
  \BibitemOpen
  \bibfield  {author} {\bibinfo {author} {\bibfnamefont {G.}~\bibnamefont {Tian}}\ and\ \bibinfo {author} {\bibfnamefont {Z.}~\bibnamefont {Gao}},\ }\bibfield  {title} {\bibinfo {title} {Frequency response analysis of active disturbance rejection based control system},\ }in\ \href {https://doi.org/10.1109/CCA.2007.4389465} {\emph {\bibinfo {booktitle} {2007 IEEE International Conference on Control Applications}}}\ (\bibinfo {year} {2007})\ pp.\ \bibinfo {pages} {1595--1599}\BibitemShut {NoStop}%
\bibitem [{\citenamefont {Vincent}\ \emph {et~al.}(2011)\citenamefont {Vincent}, \citenamefont {Morris}, \citenamefont {Usher}, \citenamefont {Gao}, \citenamefont {Zhao}, \citenamefont {Nicoletti},\ and\ \citenamefont {Zheng}}]{VINCENT201111}%
  \BibitemOpen
  \bibfield  {author} {\bibinfo {author} {\bibfnamefont {J.}~\bibnamefont {Vincent}}, \bibinfo {author} {\bibfnamefont {D.}~\bibnamefont {Morris}}, \bibinfo {author} {\bibfnamefont {N.}~\bibnamefont {Usher}}, \bibinfo {author} {\bibfnamefont {Z.}~\bibnamefont {Gao}}, \bibinfo {author} {\bibfnamefont {S.}~\bibnamefont {Zhao}}, \bibinfo {author} {\bibfnamefont {A.}~\bibnamefont {Nicoletti}},\ and\ \bibinfo {author} {\bibfnamefont {Q.}~\bibnamefont {Zheng}},\ }\bibfield  {title} {\bibinfo {title} {On active disturbance rejection based control design for superconducting {RF} cavities},\ }\href {https://doi.org/https://doi.org/10.1016/j.nima.2011.04.033} {\bibfield  {journal} {\bibinfo  {journal} {Nuclear Instruments and Methods in Physics Research Section A: Accelerators, Spectrometers, Detectors and Associated Equipment}\ }\textbf {\bibinfo {volume} {643}},\ \bibinfo {pages} {11} (\bibinfo {year} {2011})}\BibitemShut {NoStop}%
\bibitem [{\citenamefont {Qiu}\ \emph {et~al.}(2015)\citenamefont {Qiu}, \citenamefont {Michizono}, \citenamefont {Miura}, \citenamefont {Matsumoto}, \citenamefont {Omet},\ and\ \citenamefont {Sigit}}]{qiu-dob}%
  \BibitemOpen
  \bibfield  {author} {\bibinfo {author} {\bibfnamefont {F.}~\bibnamefont {Qiu}}, \bibinfo {author} {\bibfnamefont {S.}~\bibnamefont {Michizono}}, \bibinfo {author} {\bibfnamefont {T.}~\bibnamefont {Miura}}, \bibinfo {author} {\bibfnamefont {T.}~\bibnamefont {Matsumoto}}, \bibinfo {author} {\bibfnamefont {M.}~\bibnamefont {Omet}},\ and\ \bibinfo {author} {\bibfnamefont {B.~W.}\ \bibnamefont {Sigit}},\ }\bibfield  {title} {\bibinfo {title} {Application of disturbance observer-based control in low-level radio-frequency system in a compact energy recovery linac at {KEK}},\ }\href {https://doi.org/10.1103/PhysRevSTAB.18.092801} {\bibfield  {journal} {\bibinfo  {journal} {Phys. Rev. ST Accel. Beams}\ }\textbf {\bibinfo {volume} {18}},\ \bibinfo {pages} {092801} (\bibinfo {year} {2015})}\BibitemShut {NoStop}%
\bibitem [{\citenamefont {Elejaga}\ \emph {et~al.}(2024)\citenamefont {Elejaga}, \citenamefont {Jugo}, \citenamefont {Echevarria}, \citenamefont {Neumann}, \citenamefont {Ushakov},\ and\ \citenamefont {Knobloch}}]{adrc-tuner}%
  \BibitemOpen
  \bibfield  {author} {\bibinfo {author} {\bibfnamefont {A.}~\bibnamefont {Elejaga}}, \bibinfo {author} {\bibfnamefont {J.}~\bibnamefont {Jugo}}, \bibinfo {author} {\bibfnamefont {P.}~\bibnamefont {Echevarria}}, \bibinfo {author} {\bibfnamefont {A.}~\bibnamefont {Neumann}}, \bibinfo {author} {\bibfnamefont {A.}~\bibnamefont {Ushakov}},\ and\ \bibinfo {author} {\bibfnamefont {J.}~\bibnamefont {Knobloch}},\ }\bibfield  {title} {\bibinfo {title} {Experimental testing of a modified active disturbance rejection control for microphonics reduction in a 9-cell {TESLA} superconducting cavity},\ }\href {https://doi.org/10.1103/PhysRevAccelBeams.27.113501} {\bibfield  {journal} {\bibinfo  {journal} {Phys. Rev. Accel. Beams}\ }\textbf {\bibinfo {volume} {27}},\ \bibinfo {pages} {113501} (\bibinfo {year} {2024})}\BibitemShut {NoStop}%
\bibitem [{\citenamefont {Gao}(2003)}]{1242516}%
  \BibitemOpen
  \bibfield  {author} {\bibinfo {author} {\bibfnamefont {Z.}~\bibnamefont {Gao}},\ }\bibfield  {title} {\bibinfo {title} {Scaling and bandwidth-parameterization based controller tuning},\ }in\ \href {https://doi.org/10.1109/ACC.2003.1242516} {\emph {\bibinfo {booktitle} {Proceedings of the 2003 American Control Conference, 2003.}}},\ Vol.~\bibinfo {volume} {6}\ (\bibinfo {year} {2003})\ pp.\ \bibinfo {pages} {4989--4996}\BibitemShut {NoStop}%
\end{thebibliography}%

\end{document}